\documentclass[12pt]{article}
\usepackage{epsfig}
\usepackage{psfrag}
\usepackage{latexsym}
\usepackage{amssymb}
\usepackage{amsmath}
\usepackage{mathrsfs}

\def\be{\begin{equation}}
\def\ee{\end{equation}}
\def\bc{\begin{center}}
\def\ec{\end{center}}
\def\bea{\begin{eqnarray}}
\def\eea{\end{eqnarray}}
\newcommand{\om}{\omega}
\newcommand{\al}{\alpha}
\newcommand{\ba}{\begin{array}{c}}
\newcommand{\bad}{\begin{array}{ccc}}
\newcommand{\ea}{\end{array}}
\newcommand{\unity}{1\hspace{-0.15cm}1}
\def\nn{\nonumber}

\def\cC{{\cal C}}

\begin{document}
\begin{titlepage}

\vskip 2.5cm
\begin{center}
{\Large\bf A See-Saw $S_4$ model for fermion masses and mixings }
\end{center}
\vskip 0.2  cm
\vskip 0.5  cm
\begin{center}
{\large Davide Meloni }~\footnote{e-mail address: davide.meloni@physik.uni-wuerzburg.de}
\\
\vskip .2cm {\it Institut f{\"u}r Theoretische Physik und Astrophysik ,}
\\
{\it Universit{\"a}t W{\"u}rzburg, D-97074 W{\"u}rzburg, Germany}
\\
\end{center}
\vskip 0.7cm

\begin{abstract}
\noindent
We present a supersymmetric see-saw $S_4$ model giving rise to the most general neutrino mass matrix compatible with Tri-Bimaximal mixing.  We adopt the $S_4\times Z_5$  flavour symmetry, broken by suitable vacuum expectation values of a small number of flavon fields. We show that the vacuum alignment is a natural solution of the most general superpotential allowed by the flavour symmetry, without introducing any soft breaking terms. In the charged lepton sector, 
mass hierarchies are controlled by the spontaneous breaking of the flavour symmetry caused by the vevs of one doublet and one triplet flavon fields instead of using the Froggatt-Nielsen  $U(1)$ mechanism.
The next to leading order corrections to both charged lepton mass matrix and flavon vevs generate corrections to the mixing angles as large as ${\cal O}(\lambda_C^2)$. 
Applied to the quark sector, the symmetry group $S_4\times Z_5$  can give a leading order $V_{CKM}$ proportional to the identity as well as a matrix with 
${\cal O}(1)$  coefficients in the Cabibbo $2\times 2$ submatrix. Higher order corrections produce non vanishing entries in the other $V_{CKM}$ entries which are generically of ${\cal O}(\lambda_C^2)$.
\end{abstract}
\end{titlepage}
\setcounter{footnote}{0}
\vskip2truecm

\section{Introduction}
The Tri-Bimaximal structure (TBM) \cite{hps} of the neutrino mixing matrix is remarkably in agreement with the experimental results in the neutrino sector \cite{data1}. Within 1-$\sigma$ error, the values of the mixing angles can be approximated by their TBM values \cite{data}:
\bea
\tan^2 \theta_{23} = 1 \qquad \tan^2 \theta_{12} = \frac{1}{2} \qquad \sin \theta_{13} =0 \,.
\eea
The simplified structure of the mixing matrix suggested the possibility to be explained using some discrete non-abelian groups, added to the Standard Model, and containing a triplet representations to fit the number of lepton families observed in Nature. The symmetry $A_4$ \cite{TBA41,Lin:2008aj} (also in the context of Grand Unified theories \cite{TBA42}) emerged as a natural candidate because it is the smallest discrete group with triplet representation and it is sufficiently manageable to be broken differently in the charged and neutral lepton sectors, a necessary condition if we want to get a mixing matrix different from the identity.
In the context of see-saw models, it is interesting to observe that the TBM structure in $A_4$ is generally associated with a well defined relations among the complex eigenvalues of the light neutrino mass matrix:
\be
\frac{1}{m_3}=\frac{1}{m_1}-\frac{2}{m_2}~,\\
\label{sumr}
\ee
implying that most  $A_4$ models are quite predictive because of the reduced number of independent parameters. 
It should be stressed, however, that the realisation of the TBM  strongly relies on the choice of symmetry breaking pattern. In fact, 
in the neutrino sector the group $A_4$ is usually broken into a subgroup generated by the matrices 
\bea
\nn
U_{\mu-\tau}=\left(
\begin{array}{ccc}
1  &  0   & 0 \\
0 & 0  &  1 \\
0 & 1  &  0
\end{array}
\right)
\eea
and
\bea
\nn
G=\left(
\begin{array}{ccc}
-1 & 2 & 2 \\
2 & -1 & 2 \\
2 & 2 & -1 
\end{array}
\right),
\eea
which, in the basis where charged leptons are diagonal, leave invariant  the most general neutrino mass matrix diagonalized by TBM:
\bea
\label{massgen}
m_{light}=\left(
\begin{array}{ccc}
x  &  y   & y \\
y & x+v  &  y-v \\
y & y-v  &  x+v
\end{array}
\right) \,. 
\eea
It turns out that the representations of $A_4$ only contain $G$ and the invariance under $U_{\mu-\tau}$ arises accidentally, as a consequence of the specific field content of the model. From this point of view, the group $S_4$  \cite{Lam:2008sh}-\cite{bazz} arises as a natural candidate as a flavour group for neutrino mixing because one can find a suitable representation of it containing simultaneously the previous elements. 
It should also be noted that the extension of the $A_4$ symmetry from lepton to quarks seems to be complicated by the absence of doublet representations, whose use is suggested by the heaviness of the top quark, which $S_4$ possesses instead. However, the problem of reproducing small mixing angles and strong mass hierarchy in the up-sector at the same time can only be  partially alleviated by the bidimensional representations of $S_4$ because 
one or more fine tunings between the relevant Yukawas are invoked to correctly reproduce some of the $m_{up}/m_{down}$ quark mass ratios.
In this paper, we build a constrained see-saw $S_4$-based model for fermion masses and mixing which, compared
to models already existent in the literature, realises the most general neutrino mass matrix diagonalized by TBM at leading order (LO). 
This is obtained allowing the right-handed neutrinos to couple to singlet, doublet and triplet flavon fields. The light neutrino masses depend on six complex Yukawa parameters and the typical $A_4$ sum rule of eq.(\ref{sumr}) does not hold, leaving the model less predictive but more manageable. 
The mass hierarchy among charged leptons is obtained breaking the $S_4$ symmetry by the vevs of a doublet and triplet flavon fields, without invoking any Froggatt-Nielsen $U(1)$ symmetry. The unwanted couplings are forbidden imposing an additional $Z_5$ symmetry to the model. The resulting $S_4 \times Z_5$ symmetry  is minimal from the point of view of the flavour symmetry and field content. We extend the $S_4 \times Z_5$ symmetry to the quark sector including the left-handed components into  triplet representations (and not into doublets, as usually done); we show that, even using a rigid structure like that proposed in this paper, some of the relevant features of the quark sector, like a good leading order $V_{CKM}$ and quark mass ratios, can still be accounted for.
The paper is organized as follows: in Sect.\ref{structure}, we discuss the relevant feature of the $S_4$ symmetry and the structure of the model, presenting leading order results on neutrino as well as charged lepton mass matrices; in Sect.\ref{nextto} we compute the next to leading order corrections (NLO) to the vacuum alignment and the relevant higher order operators, both responsible for deviations from TBM mixing; in Sect.\ref{pheno} we discuss some phenomenological results obtained from our model with all Yakawas constrained to be ${\cal O}(1)$ and we also show
that the whole model allows for acceptable leptogenesis parameters. Sect.\ref{quarks} is devoted to the quark sector whereas in Sect.\ref{concl} we draw our conclusions.

\section{The structure of the model}
\label{structure}
We introduce here the structure of the model which leads to  TBM in first approximation.  We recall that $S_4$, the permutation group of  4 objects, can be generated by the two elements
$S$ and $T$ obeying the relations (a "presentation" of the group):
\begin{equation}
\label{4} S^4=T^3=1,~~~ST^2S=T \,.
\end{equation}
The action of the generators $S$ and $T$ can be assigned as follows:
\bea
(1234) &\rightarrow^S& (2341) \nn \cr
(1234) &\rightarrow^T& (2314) \nn
\eea
and the 24 elements of the group, belonging to 5 conjugate classes, are:
\begin{eqnarray*}
&&{\cal C}_1:1\\
&&{\cal C}_2:\;S^2=(3412),\;TS^2T^2=(4321),\;S^2TS^2T^2=(2143)\\
&&{\cal C}_3:\;T,\;T^2=(3124),\;S^2T=(1423),\;S^2T^2=(2431),\;STST^2=(4132)\\
&&~~~~~~STS=(4213),\;TS^2=(4132),\;T^2S^2=(1342)\\
&&{\cal
C}_4:\;ST^2=(1243),\;T^2S=(4231),\;TST=(1432)\\
&&~~~~~~TSTS^2=(3214),\;STS^2=(1324),\;S^2TS=(2134)\\
&&{\cal
C}_5:\;S,\;TST^2=(2413),\;ST=(3142),\;\;TS=(3421),\;S^3=(4123),\;S^3T^2=(4312)
\end{eqnarray*}
The inequivalent irreducible representations of $S_4$ are $1_1$, $1_2$, $2$ and 3. It is immediate to see that one-dimensional unitary representations are given by:
\be
\begin{array}{lll}
1_1: &S=1&T=1\\
1_2: &S=-1&T=1
\label{s$S_4$}
\end{array}
\ee
while the two-dimensional unitary representation, in a basis
where the element $T$ is diagonal, is given by:
\be
T=\left(
\begin{array}{ccc}
\omega&0\\
0&\omega^2
\end{array}
\right),~~~~~~~~~~~~~~~~
S= 
\left(
\begin{array}{ccc}
0&1\cr
1&0
\end{array}
\right)~.
\label{STbid}
\ee
Finally, the three-dimensional unitary representation is as follows:
\bea
3_1: T&=&\left(
\begin{array}{ccc}
1&0&0\\
0&\omega^2&0\\
0&0&\omega
\end{array}
\right),~~~~~~~~~~~~~~~~
S=\frac{1}{3}
\left(
\begin{array}{ccc}
-1&2\omega&2\omega^2\cr
2\omega&2\omega^2&-1\cr
2\omega^2&-1&2\omega
\end{array}
\right)~,
\label{ST}
\eea
where $\omega=e^{2\pi i/3}=(-1+\sqrt{3})/2$, whereas in the $3_2$ representation the generator $T$ is the same but $S$ is the opposite.
It is useful to remind the product rules between the group representations:
\begin{eqnarray}
\nonumber&&1_1\otimes \xi=\xi \nn \\
&&1_2\otimes 1_2=1_1,~~~~1_2\otimes2=2,~~~~1_2\otimes 3_i=3_j \\
\nonumber&&2\otimes2=1_1\oplus1_2\oplus2,~~~~2\otimes3_i=3_1\oplus3_2,~~~~3_i\otimes3_i=1_1\oplus2\oplus3_1\oplus3_2,\\
\label{6}&&3_1\otimes3_2=1_2\oplus2\oplus3_2\oplus3_2, \nn
\end{eqnarray}
where the indeces $i,j=1,2$, with $i\ne j$ and $\xi$ indicates any other representation. The Clebsch-Gordan coefficients in the basis presented above are reported in Appendix A.

The  general neutrino mass matrix of eq.(\ref{massgen}) can be obtained in the framework of the see-saw mechanism, 
\bea
\label{seesaw}
m_{light}= -m_D^T\, m_M^{-1}\, m_D~,
\eea 
where both Majorana ($m_M$) and Dirac ($m_D$) mass matrices are needed. 
They are derived from the most general lagrangian invariant under $S_4\times Z_5$
and containing fields in any of the $S_4$ representations, which are singlets, doublets and  triplets. The group 
$S_4$ is broken by means of suitable vev's of Standard Model singlet fields (flavons), whose alignments  have to guarantee the correct entries of the Majorana and Dirac mass matrices. Group theoretical considerations help in understanding the pattern of symmetry breaking needed to generate the wanted matrix; in the representation of App.A, the elements $S^2$ and $(TSTS^2)$ leave invariant the $m_{light}$ of eq.(\ref{massgen}), that is 
\bea
\nn
S^2\,m_{light}\,S^2   = m_{light}\\
(TSTS^2)\,m_{light}\,(TSTS^2)   = m_{light}\,.
\nn
\eea
This means that the vevs of the flavon fields in the neutrino sector have to be invariant under the subgroup generated by them. This strong condition for the flavon alignment is realized, for the triplet representation, by the field configuration
\bea
\label{vevS}
\langle \varphi_S \rangle = v_S\,(1,1,1)\,.
\eea 
For the bidimensional representation, the same matrices should leave the vev of a doublet field invariant; if we choose
\bea
\langle \Delta \rangle= v_\Delta\,(1,1)
\label{vevdelta}
\eea
these correspond to the first matrix of the groups ${\cal C}_{4,5}$ in the doublet representation.  Therefore, we have found a scalar field configuration which remains invariant under the action of the same matrices that leave $m_{light}$ invariant. 
In addition to the previous fields, we also include in the model  a singlet flavon field $\xi$, with a non-vanishing vev. 
To realize the classical see-saw mechanism, we need  to introduce right-handed neutrinos, which we assume to transform as a  triplet representation of $S_4$. To avoid large fine-tuning from terms of the form $M \nu^c \nu^c$, like those discussed in \cite{Lin:2008aj}, we properly tune the $\nu^c$ charge under $Z_5$\footnote{Similar considerations, but using two discrete groups, have been discussed in \cite{bm}.}; whatever this $Z_5$ charge is, if we want the singlet, doublet and triplet flavons to contribute   to the neutrino mass matrix 
their charges have to be the same. We also attribute a $Z_5$  charge to the $h_u$ higgs boson in order to avoid the Weinberg operator $O_5=\ell h_u \ell h_u$ at the leading order \cite{Weinberg:1979sa}, where $\ell$ is a triplet field of the $SU(2)$ Standard Model lepton doublets. This also forbids a leading order Dirac mass term of the form 
$(\nu^c \ell)\,h_u$, which is then generated at the ${\cal O}(1/\Lambda)$ through couplings with the same flavon fields $\xi$, $\Delta$ and $\varphi_S$.
The $Z_5$ charges for the matter fields as well as for flavons and driving fields (discussed later) are reported in  Tab.\ref{transform}, where 
we used the symbol $\omega= e^{2\pi i/5}$ to indicate the $Z_5$ unit-charge. 

\begin{table}[h!]
\centering
\begin{tabular}{|c||c|c| c|c|c||c|c|c|c|c|c|c||c|c|c|c|c|}
\hline
{\tt Field}& $\nu^c$ &$\ell$ & $e^c$ & $\mu^c$ & $\tau^c$ & $h_d$ & $h_u$& 
$\varphi_T$ & $\eta$ & $\Delta$ & $\varphi_S$ & $\xi$ & $\varphi_0^T$  & $\varphi_0^S$ & $\Delta_0$ & $\rho_0$\\
\hline
$S_4$ & $3$ &$3_1$ & $1_2$ & $1_1$ & $1_1$ & $1_1$ &$1_1$ &$3_1$ & $2$ & $2$ & $3_1$ &  $1_1$ &  $3_1$ & $3_1$ & $2$ & $1_1$\\
\hline
$Z_5$ & $\omega^2$ & 1 & $\omega^3$ & $\omega^2$ & $\omega$ & 1 & $\omega^2$&
$\omega^4$ & $\omega^4$  & $\omega$ & $\omega$ & $\omega$ &  $\omega^2$ & $\omega^3$ & $\omega^3$ & $\omega^2$\\
\hline
$U(1)_R$ & $1$ & $1$ & $1$ & $1$ & $1$ & $0$ & $0$& $0$  & $0$ & $0$ & $0$ & $0$ & $2$ & $2$ & $2$ & $2$\\
\hline
\end{tabular}
\caption{\it Transformation properties of leptons, electroweak Higgs doublets and flavons under $S_4 \times
Z_5$ and $U(1)_R$~.}
\label{transform}
\end{table}
The lagrangian in the neutrino sector is then as follows:
\bea
\label{lagneu}
W_\nu = \frac{1}{\Lambda} \nu^c \ell \,h_u \,(y_{\nu_1}\,\varphi_S+y_{\nu_2}\,\Delta+y_{\nu_3}\,\xi)+\nu^c\nu^c\, (a\,\xi+b \,\varphi_S+c \,\Delta)\,,
\eea
where $a$, $b$, $c$ and $y_{\nu_i}$ are complex Yukawa couplings.
The Dirac mass matrix is obtained from the first term in eq.(\ref{lagneu}) and it is given by:
\bea
\label{mdir}
m_D= \frac{v_u}{\Lambda}\left(
\begin{array}{ccc}
2\,y_{\nu_1}v_S+y_{\nu_3}\,u  & y_{\nu_2}\,v_\Delta-y_{\nu_1}\,v_S     & y_{\nu_2}\,v_\Delta-y_{\nu_1}\,v_S   \\
y_{\nu_2}\,v_\Delta-y_{\nu_1} \,v_S & 2\,y_{\nu_1}\,v_S +y_{\nu_2}\,v_\Delta  & y_{\nu_3}\,u-y_{\nu_1}\,v_S   \\
y_{\nu_2}\,v_\Delta-y_{\nu_1}\,v_S &  y_{\nu_3}\,u-y_{\nu_1} \,v_S &  2\,y_{\nu_1}\,v_S +y_{\nu_2}\,v_\Delta
\end{array}
\right)
\eea
where $v_u$ is the vacuum expectation value of the higgs field $h_u$.
The other terms give the Majorana mass matrix:
\bea
\label{mmleading}
m_M=\left(
\begin{array}{ccc}
a\,u+2\,b\,v_S  &  -b\,v_S+c\, v_\Delta   & -b\,v_S +c\, v_\Delta \\
-b\,v_S+c\, v_\Delta  & 2\,b\,v_S +c\, v_\Delta  &  a\,u-b\,v_S \\
-b\,v_S +c\, v_\Delta & a\,u-b\,v_S  &  2\,b\,v_S+c\, v_\Delta 
\end{array}
\right)
\eea
whose eigenvalues are:
\bea
\nn
M_1 &=& a\,u+3\,b\,v_S-c\,v_\Delta \\
M_2 &=& a\,u+2\,c\,v_\Delta \label{majeigen}\\
\nn 
M_3 &=& -a\,u+3\,b\,v_S+c\,v_\Delta \,.
\eea
Using eq.(\ref{seesaw}), we can derive the light neutrino mass matrix, diagonalized by tri-bimaximal mixing, whose eigenvalues are:
\bea
\nn 
m_1&=&-\left(\frac{v_u}{\Lambda}\right)^2\,\frac{(3\,y_{\nu_1}\,v_S-y_{\nu_2}\,v_\Delta+y_{\nu_3}\,u)^2}{a\,u+3\,b\,v_S-c\,v_\Delta} \\
m_2 &=& -\left(\frac{v_u}{\Lambda}\right)^2\,\frac{(2\,y_{\nu_2}\,v_\Delta+y_{\nu_3}\,u)^2}{a\,u+2\,c\,v_\Delta }\\
m_3&=& \left(\frac{v_u}{\Lambda}\right)^2\,\frac{(3\,y_{\nu_1}\,v_S+y_{\nu_2}\,v_\Delta-y_{\nu_3}\,u)^2}{a\,u-3\,b\,v_S-c\,v_\Delta}\, . \nn
\eea
We see that the neutrino masses depend on six unrelated complex Yukawa parameters, which offer more freedom to tune mass differences and then recover the phenomenology associated to neutrino oscillation. Notice also that no sum rules can be found in this case among complex eigenvalues.
\subsection{Charged leptons}
It could be easier to work in a basis where the charged lepton mass matrix is diagonal. In order to understand how this naturally arises in an $S_4$-based model, we observe that a generic diagonal matrix $m^D_l$ (with diagonal entries different to each other) is left invariant under the action of an element $A$ of $S_4$ only if such an element is itself diagonal, with different phase factors at each diagonal entry, as it can be understood requiring that the relation
\bea
\label{vevlept}
A^\dagger\, m^{D\dagger}_l\,m^{D}_l\,A =  m^{D\dagger}_l\,m^{D}_l
\eea
is satisfied. The generator $T$, for example, is such an appropriate matrix.  
Since the charged lepton matrix is generated after spontaneous symmetry breaking, one could choose flavon fields with vevs invariant under the action of $T$. 
Instead, we prefer the choice
\bea
\langle \varphi_T \rangle= v_T\,(0,1,0)
\eea 
which, even breaking completely the group $S_4$, not only guarantees the diagonal form of the mass matrix but also generates the hierarchy among lepton families without introducing any addition $U(1)_{FN}$ symmetry\footnote{This works in the same way as described in \cite{Lin:2008aj} noticing that, like in the case of $A_4$, $(0,1,0)^2=(0,0,1)$ and $(0,1,0)^3=(1,0,0)$.}.
In the same way,  we choose the vev for the doublet flavon $\eta$ as:
\bea
\langle \eta \rangle= v_\eta\,(0,1)\, ;
\eea 
the corresponding lagrangian in the charged lepton sector is as follows:
\bea
{\cal L}&=&\frac{y_\tau}{\Lambda} \tau^c (\ell \varphi_T) \, h_d +\nn \\
&&\frac{y_{\mu_1}}{\Lambda^2} \mu^c (\ell\, \varphi_T \,\varphi_T) \, h_d +
\frac{y_{\mu_2}}{\Lambda^2} \mu^c \ell\, (\eta\,\varphi_T) \, h_d + \nn \\
&&\label{oplept} \frac{y_{e_1}}{\Lambda^3} e^c \,\ell\,[\varphi_T (\varphi_T\varphi_T)_2]_{3_2} \, h_d +
\frac{y_{e_2}}{\Lambda^3} e^c \,\ell\,[\varphi_T (\varphi_T\varphi_T)_{3_1}]_{3_2} \, h_d +\\
&&\nn
\frac{y_{e_3}}{\Lambda^3} e^c \,\ell\, [\eta\, (\varphi_T\varphi_T)_{3_1}] _{3_2}\, h_d 
+ \frac{y_{e_4}}{\Lambda^3} e^c \,\ell\, [\varphi_T \, (\eta\eta)_{2}] _{3_2}\, h_d +\\
&&\nn
\frac{y_{e_5}}{\Lambda^2} e^c \,\ell\,  (\Delta \varphi_S)_{3_2}\, h_d\,.
\eea
It is interesting to observe that the last term in the lagrangian would be the dominant one and would drive the electron mass to a too large value\footnote{This is because this term explicitly breaks the residual symmetries needed to generate the correct hierarchies between the charged lepton masses.}. However, the leading order structures of the vacua in eqs.(\ref{vevS})-(\ref{vevdelta}) prevent this term to appear. We will show later that this is not the case when the next to leading order corrections to the flavon alignment are taken into account. 
After symmetry breaking, the mass matrix has the form:
\be
m_\ell=\frac{v_d\,v_T}{\Lambda}\left(
\begin{array}{ccc}\frac{1}{\Lambda^2}\left[v_T^2\,(y_{e_1}+2\,y_{e_2})-2\,v_T\,v_\eta\,y_{e_3}+v^2_\eta\,y_{e_4}\right]& 0& 0\\
0& \frac{1}{\Lambda}\left(2\, y_{\mu_1} \,v_T+y_{\mu_2} \, v_\eta\right)& 0\\
0& 0&  y_\tau 
\end{array}
\right)~~~,
\label{mlept}
\ee
where $v_d=\langle h_d \rangle$.
To estimate the order of magnitude of $v_T$ and $v_\eta$, we can use the experimental informations on the ratios of lepton masses. Assuming that the combinations of the $y$ coefficients are all of ${\cal O}(1)$, one obtains: 
\bea
\left(\frac{m_\mu}{m_\tau}\right)&\sim&2\,\varepsilon_T+\varepsilon_\eta \simeq 0.06 \nn \\
\left(\frac{m_e}{m_\tau}\right)&\sim&3\,\varepsilon_T^2 -2\,\varepsilon_T\,\varepsilon_\eta+\varepsilon^2_\eta \simeq 0.0003 \nn 
\eea
 where we introduced the small quantities 
\bea
\nn
\varepsilon_T=v_T/\Lambda \qquad \varepsilon_\eta=v_\eta/\Lambda\, .
\eea
These relations are satisfied for:
\bea
\label{solvev}
(|\varepsilon_T|,|\varepsilon_\eta|)&\sim&(0.017, 0.029).
\eea
so that we can roughly assume that both $\varepsilon_T$ and $\varepsilon_\eta$ are of the same order of magnitude, $\varepsilon\sim{\cal O}(\lambda_C^2)$.

\subsection{Superpotential and vacuum alignment}
\label{minim}
The most general driving superpotential $w_d$ invariant under $S_4 \times Z_5$ with $R=2$ is given by
\bea
\label{phis}
\nn 
w_d&=& g_1\, (\varphi_0^S \varphi_S\varphi_S)+
g_2 \, (\varphi_0^S \varphi_S)\,\xi+
g_3 \,\varphi_0^S (\varphi_S\Delta)+ \nn \\
&&
g_4 \,\Delta_0 (\Delta\Delta ) 
\label{wd2}
+ g_5\, \Delta_0 (\varphi_S \varphi_S)+ g_6 \,\Delta_0 \Delta \xi + 
\\
\nn 
&& 
h_1 \,\varphi_0^T (\varphi_T\varphi_T)+
h_2 \, \varphi_0^T (\eta\varphi_T) + r_1\, \rho_0 (\varphi_T\varphi_T)
 + r_2 \,\rho_0 (\eta\eta) \\ 
&=& w_d^{LO}(\varphi_0^S,\Delta_0) + w_d^{LO}(\varphi_0^T,\rho_0)  \nn
\eea
where all possible contractions among flavon fields are understood.
The equations which fix the components of the vevs of the various flavon fields are obtained solving a system of equations obtained deriving  $w_d$ with respect to the component of the driving fields. The charge assignment reported in Tab.\ref{transform} allows to separate the equations into two sets of independent relations among the fields appearing in the neutrino sector and in the charged one, respectively. For the latter we have:

\bea
\frac{\partial w_d}{\partial \varphi^T_{01}}&=&2 h_1({\varphi_T}^2_1-{\varphi_T}_2\,{\varphi_T}_3)+
h_2\,({\eta_1\,\varphi_{T_2}+\eta_2\,\varphi_{T_3}})=0\nn\\
\frac{\partial w_d}{\partial \varphi^T_{02}}&=&2 h_1({\varphi_T}^2_2-{\varphi_T}_1\,{\varphi_T}_3)+
h_2\,({\eta_1\,\varphi_{T_1}+\eta_2\,\varphi_{T_2}})=0 \label{sys:t} \\
\frac{\partial w_d}{\partial \varphi^T_{02}}&=&2 h_1({\varphi_T}^2_3-{\varphi_T}_1\,{\varphi_T}_2)+
h_2\,({\eta_1\,\varphi_{T_3}+\eta_2\,\varphi_{T_1}})=0\nn \\
\frac{\partial w_d}{\partial \rho_0}&=&r_1\,({\varphi_T}^2_1+2\,{\varphi_T}_2\,{\varphi_T}_3)+
2\,r_2\,\eta_1\,\eta_2\,, \nn
\eea
whose solutions are:
\be
\langle \eta \rangle =v_\eta\,(0, 1),~~~~~~~\langle \varphi_T \rangle =v_T\,(0, 1,0)~,~~~~~~~v_\eta=-2\left(\frac{h_1}{h_2}\right)\,v_T\,.
\label{solT}
\ee
It is important to observe that the last equation is crucial to avoid another solution of the form 
$\langle \eta \rangle =(1, -1), \langle \varphi_T \rangle =(1,1,1)$. We also recover the relation between the vevs $v_T$ and $v_\eta$ of eq.(\ref{solvev}).
In the neutrino sector, the set of equations read as follows:
\bea
\frac{\partial w_d}{\partial \varphi^S_{01}}&=&2 g_1({\varphi_S}^2_1-{\varphi_S}_2\,{\varphi_S}_3)+
g_2\,\xi\,{\varphi_S}_1+g_3\,(\Delta_1\,{\varphi_S}_2+\Delta_2\,{\varphi_S}_3)=0\nn\\
\frac{\partial w_d}{\partial \varphi^S_{02}}&=&2 g_1({\varphi_S}^2_2-{\varphi_S}_1\,{\varphi_S}_3)+
g_2\,\xi\,{\varphi_S}_3+g_3\,(\Delta_1\,{\varphi_S}_1+\Delta_2\,{\varphi_S}_2)=0 \nn  \\
\label{sys:s}
\frac{\partial w_d}{\partial \varphi^S_{03}}&=&2 g_1({\varphi_S}^2_3-{\varphi_S}_1\,{\varphi_S}_2)+
g_2\,\xi\,{\varphi_S}_2+g_3\,(\Delta_1\,{\varphi_S}_3+\Delta_2\,{\varphi_S}_1)=0 \\
\frac{\partial w_d}{\partial \Delta_{01}}&=&
g_4\,\Delta_1^2+g_5\,({\varphi_S}^2_3+2\,{\varphi_S}_1\,{\varphi_S}_2)+g_6\,\Delta_2\,\xi=0 \nn \\
\frac{\partial w_d}{\partial \Delta_{02}}&=&
g_4\,\Delta_2^2+g_5\,({\varphi_S}^2_2+2\,{\varphi_S}_1\,{\varphi_S}_3)+g_6\,\Delta_1\,\xi=0 \,.\nn
\eea
The system is solved by:
\bea
\label{solS}
\langle \xi \rangle =u,~~~~~~~\langle \Delta \rangle =v_\Delta\,(1, 1),~~~~~~~\langle \varphi_S \rangle =v_S\,(1, 1,1)\,, 
\eea 
with the additional relations
\bea
v_\Delta&=&-\frac{g_2\,u}{2\,g_3} \\
v_S^2&=& \left(\frac{2 g_2\, g_3\,g_6-g_2^2\,g_4} {12 g_5\,g_3^2}\right)\,u^2
\label{vsu}
\eea
and $u$ undetermined.
We have explicitly checked that the solutions of the vacuum alignment equations are unique; in fact, requiring for the generic field $\Phi^{LO}$ to be shifted as $\Phi^{LO}+\delta\Phi$, we found that the components  $\delta\Phi$ are all in the same directions of the corresponding $\Phi^{LO}$; thus, we do not need to introduce any soft term to drive the superportential into the wanted minimum.

\section{Next to leading order}
\label{nextto}
The discussion of the corrections to the previous results starts with a study of the next to leading order structure of the vacuum alignments of the flavon fields. It will turn out that such corrections will be enough to guarantee deviation from TBM at a level compatible with the recent experimental results.

\subsection{Corrections to the vacuum alignment}

The next to leading order terms mix the charged lepton and neutrino sectors in a non-trivial way. 
We want to find perturbations of (\ref{solT},\ref{solS}) of the form
\bea
\langle \varphi_S \rangle &=&(v_S+\delta v_{S_1}, v_S+\delta v_{S_2},v_S+\delta v_{S_3}) \nn \\
\langle \varphi_T \rangle &=&(\delta v_{T_1}, v_T+\delta v_{T_2},v_T+\delta v_{T_3})\nn\\
\langle \Delta \rangle &=& (v_\Delta+ \delta v_{\Delta_1}, v_\Delta+ \delta v_{\Delta_2} )  \\
\langle \eta \rangle &=& ( \delta v_{\eta_1}, v_\eta+ \delta v_{\eta_2} ) \nn
\\ \nn 
\langle \xi \rangle &=& u+\delta u \,.
\eea  
On a general ground, we have eleven unknowns but only nine equations (six from the two triplets, two from the doublet and one from the singlet); then, we expect two of the previous shifts to remain unconstrained.
To better understand the output of such an analysis, we study the $(\varphi^S_0, \Delta_0)$ and $(\varphi^T_0, \rho_0)$ sectors separately.
\subsubsection{The $ (\varphi_0^S, \Delta_0)$ sector} 
This sector is responsible for the alignment of the fields $\varphi_S$ and $\Delta$.
At order ${\cal O}(1/\Lambda)$, the three-field terms entering the superpotential are combination of $\varphi_T$ and $\eta$; collecting these terms, we have:
\bea
\label{nlophis}
\delta w_d(\varphi^S_0) = \Sigma_{i=1}^{3} \frac{s_i}{\Lambda}\, \varphi_0^S\,(\varphi_T\varphi_T\varphi_T)_i + 
\Sigma_{i=4}^{5} \frac{s_i}{\Lambda}\, \varphi_0^S\,(\varphi_T\eta\eta)_i +
\frac{s_6}{\Lambda}\, \varphi_0^S\,(\varphi_T\varphi_T\eta)_i 
\eea
\bea
\label{nlodelta}
\delta w_d(\Delta_0) = \Sigma_{i=1}^{2} \frac{\delta_i}{\Lambda}\, \Delta_0\,(\eta\eta\eta)_i + 
 \Sigma_{i=3}^{4} \frac{\delta_i}{\Lambda}\, \Delta_0\,(\varphi_T\varphi_T\eta)_i +
\frac{\delta_5}{\Lambda}\, \Delta_0\,(\varphi_T\varphi_T\varphi_T)_i  
\eea
where, in both cases,  the index $i$ of the trilinear terms represents different $S_4$ contractions. 
These NLO corrections have to be evaluated with the leading order vev's in eq.(\ref{solT}), so that they do not contain any of the unknowns under investigation. In particular, the structure of the vacua (\ref{solT}) produces 
\bea
\nn
\delta w_d(\Delta^0) = 0\, ;
\eea
and the total LO+NLO part of the superpotential responsible for the alignment of $\varphi_S$ and $\Delta$ is given by:
\bea\nn
w_d(\varphi_0^S,\Delta_0)=w_d^{LO}(\varphi_0^S,\Delta_0)+\delta w_d(\varphi^S_0)\,.
\eea
Symmetry arguments allow to understand the structure of the solutions, whose detailed expressions can be  found by explicitly solving the system of equations according to the procedure of Sect.(\ref{minim}). In fact, we see that, after symmetry breaking,  the non-vanishing terms in $\delta w_d(\varphi^S_0)$ are proportional to the vector $(1,0,0)$, which is left invariant by the $S_4$ element $(TSTS^2)$; this has a $2 \leftrightarrow 3$ symmetry which forces the new vev of $\varphi_S$ to have the form  $(v_S+\delta v_{S_1}, v_S+\delta v_{S_2},v_S+\delta v_{S_2})$. At the same time, the shifts $\delta v_{\Delta_i} $ should all be equal because of the vanishing NLO term $\delta w_d(\Delta^0)$. An explicit computation confirms these speculations and we can write the new vacua in the following form: 
\bea
\label{phisnlo}
\varphi_S=v_S\,\left(
\begin{array}{c}
1 + A_S\,\left(\frac{\varepsilon}{\varepsilon'}\right)^2\, \varepsilon \\ \\
1 + B_S\, \left(\frac{\varepsilon}{\varepsilon'}\right)^2\,\varepsilon  \\  \\
1 + B_S\, \left(\frac{\varepsilon}{\varepsilon'}\right)^2\,\varepsilon 
\end{array}
\right) \qquad
\Delta=v_\Delta\,\left(
\begin{array}{c}
1 + A_\Delta\,\left(\frac{\varepsilon}{\varepsilon'}\right)^2\, \varepsilon \\ \\
1 + A_\Delta\,\left(\frac{\varepsilon}{\varepsilon'}\right)^2\, \varepsilon  
\end{array}\right)
\eea
where the coefficients $A_S$, $B_S$ and $A_\Delta$ are linear combinations of leading and next to leading order coefficients and, for the sake of simplicity, we have introduced the small parameter $\varepsilon'=\langle\varphi_S\rangle/\Lambda\sim\langle\Delta\rangle/\Lambda\sim\langle\xi\rangle/\Lambda$. Notice that the three $\delta v_{S_i}$ also depend on the undetermined parameter  $\delta u$ with identical coefficients, so that they can be readsorbed in the leading order result.

\subsubsection{The $ (\varphi_0^T, \rho_0)$ sector} 
This sector is responsible for the alignment of the fields $\varphi_T$ and $\eta$.
At order ${\cal O}(1/\Lambda)$, the new terms in the superpotential are combinations of $\varphi_S$, $\Delta$ and $\xi$; we have:
\bea
\label{nlophit}
\delta w_d(\varphi^T_0) &=& \Sigma_{i=1}^{3} \frac{t_i}{\Lambda}\, \varphi_0^T\,(\varphi_S\varphi_S\varphi_S)_i + 
\Sigma_{i=4}^{5} \frac{t_i}{\Lambda}\, \varphi_0^T\,(\varphi_S\Delta\Delta)_i + \\ \nn
&&\frac{t_6}{\Lambda}\, \varphi_0^T\,(\varphi_S\varphi_S\Delta) + 
\frac{t_7}{\Lambda}\, \varphi_0^T\,(\varphi_S\varphi_S) \xi + 
\frac{t_8}{\Lambda}\, \varphi_0^T\,\varphi_S\, \xi^2 + 
\frac{t_9}{\Lambda}\, \varphi_0^T\,(\varphi_S\,\Delta)\, \xi
\eea
and
\bea
\label{nlorho}
\delta w_d(\rho_0) &=& 
\frac{\rho_1}{\Lambda}\, \rho_0\,(\varphi_S\varphi_S\varphi_S)  + 
\frac{\rho_2}{\Lambda}\, \rho_0\,(\Delta\varphi_S\varphi_S) + 
\frac{\rho_3}{\Lambda}\, \rho_0\,(\varphi_S\, \varphi_S)\, \xi +\\
&&\frac{\rho_4}{\Lambda}\, \rho_0\,(\Delta\,\Delta\,\Delta) +
\frac{\rho_5}{\Lambda}\, \rho_0\,(\Delta\,\Delta)\,\xi +  \nn
\frac{\rho_5}{\Lambda}\, \rho_0\, \xi^3 \, .
\eea
In this case, the corrections to $\varphi^T_0$ do not conserve any of the $S_4$ basis elements and we expect that the components of $\langle\varphi_T\rangle$ will point to  different directions.  Also the $\langle\eta\rangle$'s vev  is not preserved at the next to leading order and, consequently, the shifts $ \delta v_{\eta_i}$ are different from zero and different to each other. We choose to treat $ \delta v_{\eta_2}$ as the second undetermined parameter, so that the new vacua can be cast in the following form:
\bea
\label{phitnlo}
\varphi_T=v_T\,\left(
\begin{array}{c}
A_T\,\left(\frac{\varepsilon'}{\varepsilon}\right)^2\, \varepsilon' \\ \\
1 + B_T\,\left(\frac{\varepsilon'}{\varepsilon}\right)^2\, \varepsilon'  \\ \\
C_T\,\left(\frac{\varepsilon'}{\varepsilon}\right)^2\, \varepsilon' 
\end{array}
\right) \qquad
\eta=v_\eta\,\left(
\begin{array}{c}
A_\eta\,\left(\frac{\varepsilon'}{\varepsilon}\right)^2\, \varepsilon' \\ \\
1 + B_\eta\, \delta v_{\eta_2}  
\end{array}\right)\,.
\eea 
Notice that a correction proportional to $ \delta v_{\eta_2}$ also appears in the second component of $\varphi_T$ in the form $\varphi_{T_2}=v_T-(h_2/2 h_1)\,\delta v_{\eta_2}+{\cal O}(\varepsilon')$ which, using the last relation in eq.(\ref{solT}), 
gives $\varphi_{T_2}=-(h_2/2 h_1)\,(v_T+\delta v_{\eta_2})+{\cal O}(\varepsilon')$, so that 
the shift $ \delta v_{\eta_2}$ can be readsorbed into a redefinition of $v_T$.

\subsection{Charged lepton mass matrix}
The next to leading order corrections to the mass matrix of eq.(\ref{mlept}) come from 
the corrections to the vacuum alignments in eqs.(\ref{phisnlo}) and (\ref{phitnlo})
and from higher dimensional operators, suppressed by a relative ${\cal O}(1/\Lambda)$ with respect to each of the terms quoted in eq.(\ref{oplept}). In the following we will study them separately.

\subsubsection{Corrections from vacuum alignment}

The main features of these corrections are related to the fact that all the vanishing entries in the matrix of eq.(\ref{mlept}) are suppressed by one additional ${\cal O}(1/\Lambda)$ factor compared to the diagonal entries, except for the first line, where the ${\cal O}(1/\Lambda^2)$ operator in eq.(\ref{oplept}) is non vanishing. As a result, we get the following charged lepton mass matrix at the NLO:

\be
m_\ell=v_d\left(
\begin{array}{ccc} a_1 \, \varepsilon^3 &a_2 \,\varepsilon\, \varepsilon^{'2}& -a_2 \,\varepsilon\,\varepsilon^{'2}\\
b_1\, \varepsilon^{'3}& b_2\,\varepsilon^2 & b_3\,\varepsilon^{'3}\\
c_1\,\frac{\varepsilon^{'3}}{\varepsilon}  & c_2\,\frac{\varepsilon^{'3}}{\varepsilon}  & c_3\,\varepsilon
\end{array}
\right) 
\label{mleptfin}
\ee
where the coefficients $a_i, b_i$ and $c_i$ can be easily reconstructed from  eqs.(\ref{oplept}) and (\ref{phitnlo}). The matrix 
$m_\ell^\dagger\, m_\ell$ can be diagonalized  by the unitary transformation 
\be
U_\ell= \left(
\begin{array}{ccc} 1 & (\frac{b_1}{b_2}\,\frac{\varepsilon^{'3}}{\varepsilon^2})^* &  (\frac{c_1}{c_3} \,\frac{\varepsilon^{'3}}{\varepsilon^2})^*  \\
-\frac{b_1}{b_2}\,\frac{\varepsilon^{'3}}{\varepsilon^2}& 1 & (\frac{c_2}{c_3}\, \frac{\varepsilon^{'3}}{\varepsilon^2})^*\\
-\frac{c_1}{c_3} \,\frac{\varepsilon^{'3}}{\varepsilon^2}& -\frac{c_2}{c_3} \,\frac{\varepsilon^{'3}}{\varepsilon^2} & 1
\end{array}
\right)~. 
\label{ue}
\ee
so that, at the lowest order in $\varepsilon'$, the coefficients of the electron row in eq.(\ref{mleptfin})
do not contribute. From the matrix in eq.(\ref{ue}), we can compute the $U_{\rm PMNS}$ mixing matrix from the lepton sector only:
\bea
&& U_{\rm PMNS}= U_\ell^\dagger\,U_{TBM} \nn 
=\\
&&\hspace{-1cm}\left(
\begin{array}{ccc} \sqrt{\frac{2}{3}} +  \frac{1}{\sqrt{6}} \,(a^*+b^*)\,\left(\frac{\varepsilon^{'3}}{\varepsilon^2}\right)^* &  
\frac{1}{\sqrt{3}} -  \frac{1}{\sqrt{3}} \,(a^*+b^*)\,\left(\frac{\varepsilon^{'3}}{\varepsilon^2}\right)^* &  \frac{1}{\sqrt{2}} \,(a^*-b^*)\,\left(\frac{\varepsilon^{'3}}{\varepsilon^2}\right)^*\\
-\sqrt{\frac{1}{6}} +  \sqrt{\frac{2}{3}}\,a\,\left(\frac{\varepsilon^{'3}}{\varepsilon^2}\right) +  \frac{1}{\sqrt{6}}\,c^*\,\left(\frac{\varepsilon^{'3}}{\varepsilon^2}\right)^*  & 
\sqrt{\frac{1}{3}} +  \frac{1}{\sqrt{3}}\,a\,\left(\frac{\varepsilon^{'3}}{\varepsilon^2}\right) -  \frac{1}{\sqrt{3}}\,c^*\,\left(\frac{\varepsilon^{'3}}{\varepsilon^2}\right)^* & 
-\sqrt{\frac{1}{2}}-c^*\,\left(\frac{\varepsilon^{'3}}{\varepsilon^2}\right)^*\\
-\sqrt{\frac{1}{6}} +  \frac{1}{\sqrt{6}}\,(2 b-c)\, \left(\frac{\varepsilon^{'3}}{\varepsilon^2}\right)   & 
\sqrt{\frac{1}{3}} +  (b+c)\,\left(\frac{\varepsilon^{'3}}{\varepsilon^2}\right)   &   
\sqrt{\frac{1}{2}} -\sqrt{\frac{1}{2}}\,c\,\left(\frac{\varepsilon^{'3}}{\varepsilon^2}\right)
\end{array}
\right) \nn \\  && \label{upmnslept}
\eea
where $a=b_1/b_2$, $b=c_1/c_3$ and $c=c_2/c_3$.
Then, any entries of $U_{\rm TBM}$ get corrected by terms of  ${\cal O}(\varepsilon'^3/\varepsilon^2)$  
which can be as large as ${\cal O}(\lambda_C^2)$ to fit the experimental data.

\subsubsection{Corrections from higher dimensional operators}
For the electron case, there are many of such  operators, coming from the following contractions:
\bea
&& \left(\Delta^3 \varphi_T\right) \,  \left(\Delta^3 \eta\right) \,\left(\varphi_S^3 \varphi_T\right) \, 
\left(\varphi_S^3 \eta\right) \,\left(\Delta^2 \varphi_T\varphi_S\right) \nn
\\ \nn  \\  
&&
\left(\Delta^2 \varphi_T\xi\right)\,\left(\Delta^2 \varphi_S\eta\right) \,\left(\Delta^2 \xi\eta\right)\,
\left(\varphi_S^2 \varphi_T\Delta\right)\,\left(\varphi_S^2 \varphi_T\xi\right) 
 \\ \nn  \\ \nn
&& 
\left(\varphi_S^2 \Delta\eta\right)\,\left(\xi^2 \varphi_T\Delta\right) \,\left(\xi^2\varphi_S\eta\right)
\,\left(\xi^2\varphi_T\varphi_S\right)
\,\left(\xi^2\Delta\varphi_S\right)\,.
\eea
It is easy to understand that they contribute to the electron row by terms of order ${\cal O}(1/\Lambda^4)$, which are unimportant because they do not sizebly modify the 
diagonalizing matrix $U_\ell$. Also, the corrections to the $\tau$ row are of ${\cal O}(1/\Lambda^3)$, that is the coefficients $b$ and $c$ appearing in the $U_{PMNS}$ are modified at the next to next to leading order (NNLO) and can be safely neglected. 
For the muon case, the following higher dimensional operators  modify the coefficient $a$ appearing in eq.(\ref{upmnslept}) at the NLO :
\bea
{\cal L^{\it NLO}}&=&\Sigma_{i=1}^{2}\frac{y_{\mu^{NLO}_i}}{\Lambda^3} \mu^c \ell\, (\varphi_S \varphi_S  \varphi_S)\,h_d +
\Sigma_{i=3}^{4}\frac{y_{\mu^{NLO}_i}}{\Lambda^3} \mu^c \ell\, (\varphi_S \Delta  \Delta)\,h_d + \\ \nn &&
\frac{y_{\mu^{NLO}_5}}{\Lambda^3} \mu^c \ell\,\varphi_S\,\xi^2 + 
\frac{y_{\mu^{NLO}_5}}{\Lambda^3} \mu^c \ell\,(\varphi_S\Delta)\,\xi\,.
\eea 
In conclusion, both types of corrections (from vacuum alignment and from higher dimensional operators) to the charged lepton mass matrix contribute to generate deviation from TBM at ${\cal O}(\lambda_C^2)$.

\subsection{Neutrinos}
Also for neutrinos, we have to take into account corrections from both vacuum alignment and higher dimensional operators. 
In particular, the NLO corrections of the vevs of the flavon fields affect both  Majorana and Dirac masses with the same pattern, as we can see from eq.(\ref{lagneu}). Different higher order corrections are generated by higher order operators. These corrections turn out to be negligible compared with the NLO and of the same order of magnitude as those induced by 
the  Weinberg operator (see later). 

\subsubsection{Corrections to the Majorana and Dirac mass matrices}
\label{subcorr}
The relevant contributions come from the NLO vacuum alignments. In particular, eq.(\ref{mmleading}) is modified only by $\langle \varphi_S \rangle$ at the NLO because the other two vevs, namely those of $\Delta$ and $\xi$ entering in the lagrangian in eq.(\ref{lagneu}), are aligned along the LO direction. However, the shifts in $\langle \varphi_S \rangle$ are not enough to introduce independent corrections to all the (six) elements of the Majorana and Dirac mass matrices; in fact, the  
explicit  structure of the NLO corrections are as follows:
\bea
\label{maysec}
\delta m_{M}&=&b\,\left(\frac{\varepsilon}{\varepsilon'}\right)^2\,\varepsilon\,v_S\,\hat S \\ \nn \\
\label{diracsec}
\delta m_{D}&=&v_u\,\left(\frac{\varepsilon}{\varepsilon'}\right)\,\varepsilon^2\,y_{\nu_1}\,\hat S
\eea
where the common matrix $\hat S$ is:
\bea
\label{hatS}
\hat S=\left(
\begin{array}{ccc}
2 \,A_S  &  -B_S        &  -B_S \\
 -B_S &   2\,B_S    &   -A_S    \\
 -B_S &   -A_S &    2\,B_S 
\end{array}
\right) \,.
\eea
With these results, we can build the light neutrino mass matrix; since the NLO matrix elements have now complicated expressions in terms of the leading order Yakawas of eq.(\ref{lagneu}) and of the correction coefficients $A_S$ and $B_S$, we prefer to summarize its structure as follows:
\bea
\label{massgenNLO}
m_{ light}=\varepsilon^{'2}\,\left(
\begin{array}{ccc}
x  &  y   & y \\
y & x+v  &  y-v \\
y & y-v  &  x+v
\end{array}
\right) + \varepsilon^3\,\left(
\begin{array}{ccc}
x'  &  y'  & y' \\
y' & z  &  z' \\
y' & z'  &  z
\end{array}\right) \,,
\eea  
where the relation $z+z'\ne x'+y'$ implies that $m_{ light}$ is not diagonalized by $U_{TBM}$. However, 
it is easy to show that the almost symmetric structure of the $\varepsilon^3$ contribution prevents  to generate corrections to the TBM values 
of $\theta_{13}$ and $\theta_{23}$  because 
the diagonalization of  $m_{light}$ is achieved by a matrix $U_{\nu}=U_{TBM}+\delta U$, where 
$\delta U$ has a vanishing last column. 
If we want to obtain  a non-vanishing $\theta_{13}$, we need 
to distinguish the elements (12) from (13) and/or (22) from (33) in eq.(\ref{hatS}). Without taking into account the corrections from the charged lepton sector, this can be accomplished only at the NNLO. 
In fact, a $Z_5$ singlet built with the  $\nu^c \nu^c$ bilinear requires a charge $\omega$ for the accompaining fields, and this can be achieved with at least three fields, which induce terms of ${\cal O}(1/\Lambda^2)$. There are more than twenty of these terms, generated by the following contractions:
\bea
&&\left(\Delta^2 \varphi_T\right) \,  \left(\Delta^2 \eta\right) \,\left(\varphi_S^2 \varphi_T\right) \, 
\left(\varphi_S^2 \eta\right) \,\left(\xi^2 \varphi_T\right) \,
\left(\xi^2 \eta\right) \nn \\ \nn  \\ \nn
&& \,\left(\varphi_T\Delta\varphi_S\right)\,\left(\varphi_T\Delta\xi\right)\,\left(\varphi_T\varphi_S\xi\right)\,
\left(\Delta\eta\varphi_S\right)\,
\left(\Delta \eta \xi\right) \, \left(\varphi_S\eta\xi\right)\,.
\eea
Similarly, the Dirac mass matrix receives corrections at the same NNLO relative to the leading terms due to the fact that the term $(\nu^c \ell)\,h_u$ has a total $Z_5=\omega^4$.

Beside the previous operators, we also have to consider effective terms of the form $l l h_u h_u$; given its charge assignment ($Z_5=\omega^4$), the Weinberg operator  arises at ${\cal O}(1/\Lambda^2)$ with the insertion of one flavon field, as for the Dirac and Majorana terms:
 
\bea
W_\nu^{\rm eff} &=\frac{1}{\Lambda^2}\,(\alpha_1\varphi_S +\alpha_2 \Delta + \alpha_3 \xi)   \left(\ell \, h_u \,\ell \, h_u\right) \,.
\eea 
After spontaneous symmetry breaking, $W_\nu^{\rm eff}$ generates terms which are of order:
\bea
\label{weinberg}
m_{W}\sim \frac{v_u^2\, \langle\Phi\rangle}{\Lambda^2} \sim \varepsilon'\, \left(\frac{v_u}{\Lambda}\right)\,v_u\,.
\eea
Compared to the NLO corrections of the Dirac mass term, eq.(\ref{diracsec}), we 
see that: 
\bea
m_{W}/\delta m_{D} \sim \frac{v_u/\Lambda}{\varepsilon} << 1\,;
\nn
\eea
then this type of operators is more suppressed with respect to the Dirac mass corrections (and even more if compared with the Majorana mass terms) and can be safely neglected.

In conclusion, from the neutrino sector only, the following mixing matrix arises:
\bea
U_\nu=
&& \left(
\begin{array}{ccc} \sqrt{\frac{2}{3}} -  \sqrt{\frac{2}{3}} \,\frac{(x'+y'-z-z')^*}{9y^*}\,\left(\frac{\varepsilon^{3}}{\varepsilon^{'2}}\right)^* &  
\frac{1}{\sqrt{3}} + 2 \,\frac{(x'+y'-z-z')}{9\sqrt{3}y}\,\left(\frac{\varepsilon^{3}}{\varepsilon^{'2}}\right)& 0\\
-\sqrt{\frac{1}{6}} -  \sqrt{\frac{2}{3}} \,\frac{(x'+y'-z-z')^*}{9y^*}\,\left(\frac{\varepsilon^{3}}{\varepsilon^{'2}}\right)^*  & 
\sqrt{\frac{1}{3}} - \,\frac{(x'+y'-z-z')}{9\sqrt{3}y}\,\left(\frac{\varepsilon^{3}}{\varepsilon^{'2}}\right)  & 
-\sqrt{\frac{1}{2}}\\
-\sqrt{\frac{1}{6}} -  \sqrt{\frac{2}{3}} \,\frac{(x'+y'-z-z')^*}{9y^*}\,\left(\frac{\varepsilon^{3}}{\varepsilon^{'2}}\right)^*  & 
\sqrt{\frac{1}{3}} - \,\frac{(x'+y'-z-z')}{9\sqrt{3}y}\,\left(\frac{\varepsilon^{3}}{\varepsilon^{'2}}\right)    &   
\sqrt{\frac{1}{2}} 
\end{array}
\right) \nn
 \\ && \label{domega}
\eea

\subsubsection{Mixing angles at the NLO}
\label{subcorr2}
From the previous discussions, it clearly appears that the NLO corrections to $\theta_{13}$ and  $\theta_{23}$ come only from  $U_\ell$ whereas $\theta_{12}$ is modified by both $U_\ell$ and $U_\nu$.
They are as follows:

\bea
s_{13}&=&|U_{e3}|=\left|\frac{1}{\sqrt{2}} \,(A-B)^*\,\left(\frac{\varepsilon^{'3}}{\varepsilon^2}\right)^*\right| \nn \\
s_{12}&=& \frac{|U_{e2}|}{\sqrt{1-|U_{e3}|^2}} =\left|\frac{1}{\sqrt{3}} -  \frac{1}{\sqrt{3}} \,(A+B)^*\, \left(\frac{\varepsilon^{'3}}{\varepsilon^2}\right)^* -\sqrt{2}\alpha \left(\frac{\varepsilon^3}{\varepsilon^{'2}}\right)\right|\\
s_{23}&=&\frac{|U_{\mu 3}|}{\sqrt{1-|U_{e3}|^2}}=\left|\frac{1}{\sqrt{2}}+C^*\,\left(\frac{\varepsilon^{'3}}{\varepsilon^2}\right)^*\right| \nn
\eea
where we used the short-hand notation
\bea
\alpha &=& -  \sqrt{\frac{2}{3}} \,\frac{(x'+y'-z-z')}{9y}\,. \nn 
\eea
In particular, the first relation can be used to put a bound on $|\varepsilon'|$; in fact, given the maximum allowed value for $\theta_{13}$ and assuming that $|A-B|\sim {\cal O}(1)$ we get:
\bea
|\varepsilon'|\lesssim \left[\sqrt{2}\,\lambda_C^4\,\theta_{13}^{\rm max}\right]^{1/3}
\eea
which is close to $(2\,\lambda_C^2)$ for $\theta_{13}^{\rm max}\sim 10^o$.

\section{A bit of phenomenology}
\label{pheno}
\begin{figure}[ht!]
\hspace{-0.3cm}
\centering
\includegraphics[width=0.5\linewidth]{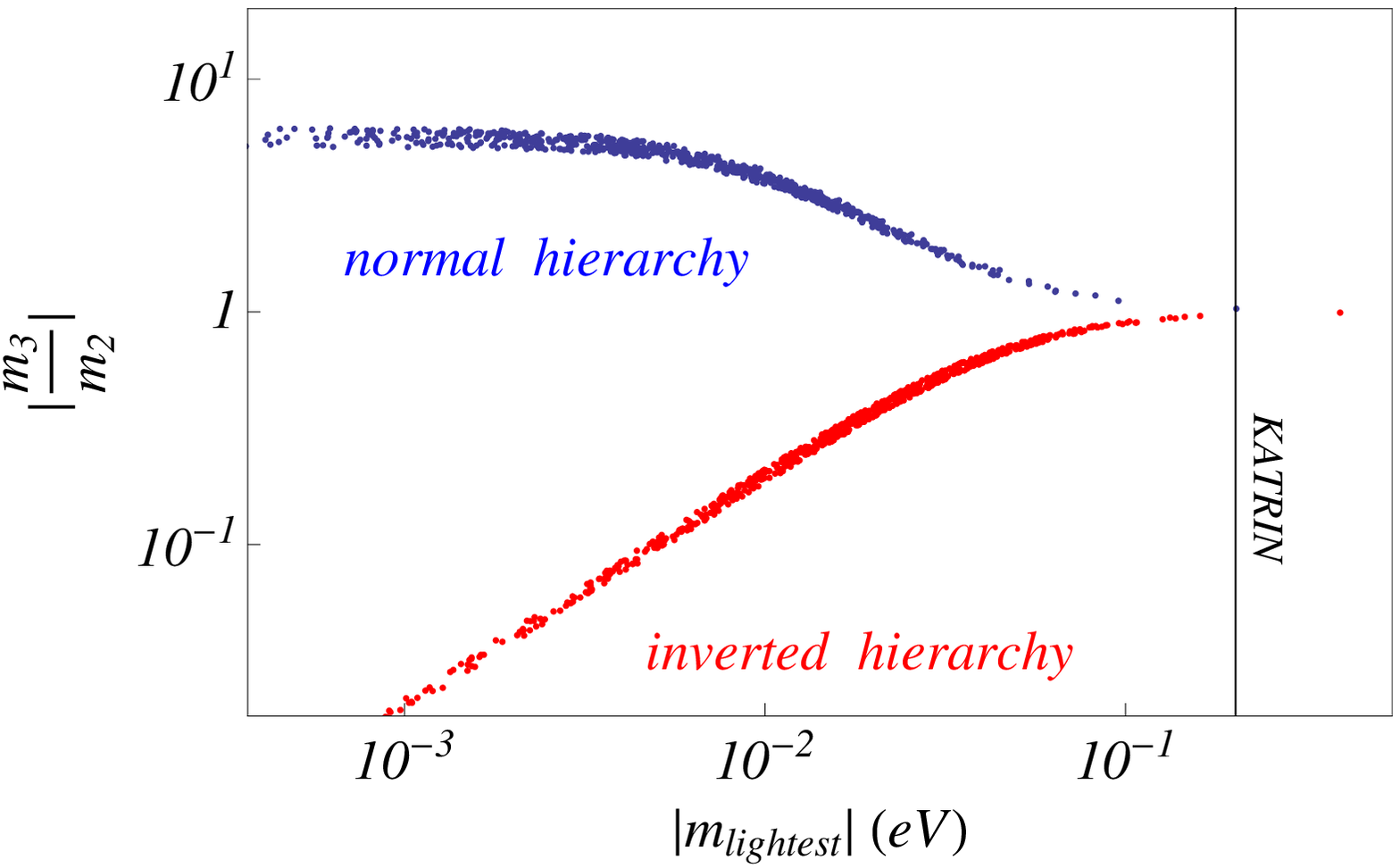} 
\includegraphics[width=0.473\linewidth]{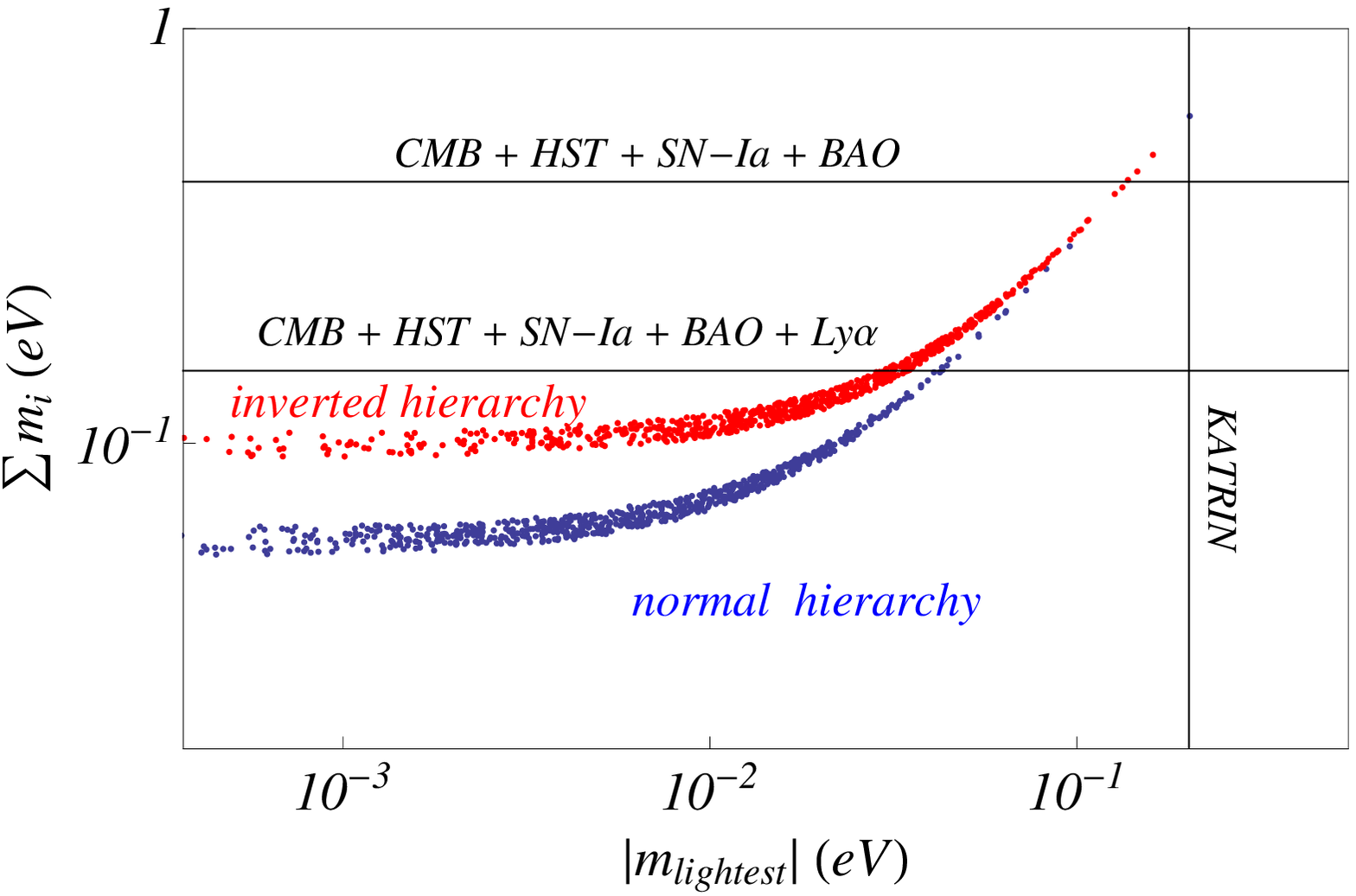}
\caption{\label{pl:masses} \it Neutrino mass spectrum and sum of neutrino masses, as predicted from the model with all Yukawa couplings of ${\cal O}(1)$ and the flavon vevs of the same order of magnitude. Left panel: behaviour of the ratio $|m_3|/|m_2|$  as a function of the lightest neutrino mass $|m_{lightest}|$, for both neutrino hierarchies. Right panel: sum of the light neutrino masses as a function of $|m_{lightest}|$. Also shown are the bounds from \cite{set1} (upper solid lines) and from  \cite{set1} +  \cite{Ly1} (lower solid lines). The vertical line in both panels is the future sensitivity  of 0.2 eV on $m_{lightest}$  from the KATRIN
experiment \cite{Osipowicz:2001sq}.}
\end{figure}

The model we have presented has a huge parameter space, made by the six complex Yukawa  couplings $y_{\nu_i}$ and  $a,b$ and $c$. It is clear, and we have checked this numerically, that the various phenomenological constraints, coming for example from the smallness of the parameter  $r=\Delta m^2_{sol}/|\Delta m^2_{atm}|$, are easily satisfied, for both type on neutrino mass hierarchies.
It seems then more interesting to ask whether the model can still give an acceptable phenomenology in some particular cases, like for example allowing all the Yukawa couplings to be of ${\cal O}(1)$ and the flavon vevs $\langle\varphi_S\rangle, \langle\Delta\rangle$ and $\langle\xi\rangle$ to be of the same order of magnitude\footnote{In the numerical simulation, this means that we allow the absolute values of the Yukawas to be in the interval $[1/2,3/2]$, while no restriction whatsoever has been imposed on the flavon vevs a part from being all equal within a factor of 10; also the large scale $\Lambda$ is left free.}. In this case, we do not expect any huge hierarchies among the heavy neutrinos, see eq.(\ref{majeigen}). 
Our numerical study aims to predict some interesting physical quantities only imposing the 
following 3-$\sigma$ experimental constraints \cite{MaltoniIndication}:
\bea
\Delta m^2_{sol} &>& 0 \nn \\
|\Delta m^2_{atm}| &=&  2.41 \pm 0.34\times 10^{-3}\, eV^2 \label{condnorm} \\
r^{\rm exp}&=& 0.032 \pm 0.006\nn~,
\eea
also taking $|m_i| \lesssim 0.5~eV$. 
The resulting spectrum of the light  neutrino masses and their sum is shown in Fig.(\ref{pl:masses}).
In the left panel, we chose to plot the ratio $|m_3|/|m_2|$ as a function of the lightest neutrino mass $|m_{lightest}|$, which is $|m_1|(|m_3|)$ for the normal (inverted) hierarchy. We clearly see that the largest hierarchies, obtained at the smallest allowed $|m_{lightest}|$, are at the level of ${\cal O}(10)$ for the normal ordering and ${\cal O}(10^{-2})$ for the inverted one. The ratio tends to a degenerate spectrum in both cases for $|m_{lightest}|\sim 10^{-1}$ eV which, however, is disfavoured in our model. In the right panel we show the sum of the light neutrino masses as a function of the lightest neutrino mass $m_{lightest}$.
The vertical line denotes the future sensitivity  of 0.2 eV on of  $|m_{lightest}|$  from the KATRIN
experiment \cite{Osipowicz:2001sq}, and the horizontal lines are the cosmological bounds
\cite{Fogli:2008cx} at $0.60$ eV, obtained combining the data from ref.\cite{set1}, and 
at $0.19$ eV, corresponding to
all the previous data combined to the small scale primordial
spectrum from Lyman-alpha (Ly$\alpha$) forest clouds \cite{Ly1}. 
We see that  our model predicts $\Sigma \,m_i$ too similar for both hierarchies to be distinguished 
using the current cosmological information on the sum of the
neutrino masses; however, such a  discrimination could be possible if some improvements on these bounds 
would be achieved in the near future.

\begin{figure}[h!]
\hspace{-0.3cm}
\centering
\includegraphics[width=0.6\linewidth]{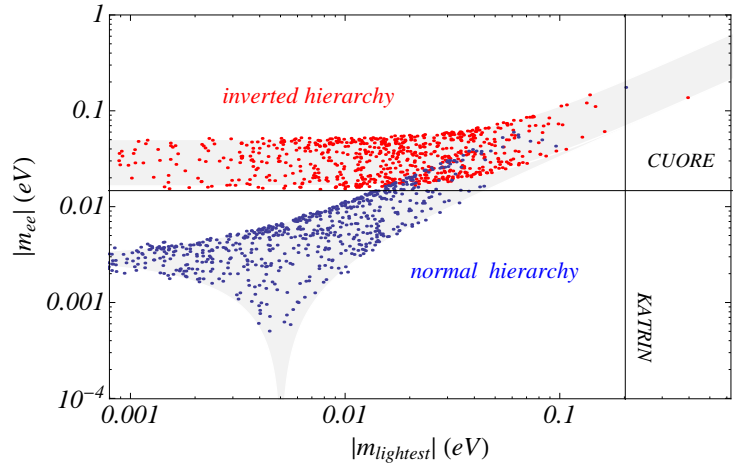}
\caption{\label{mee} \it $|m_{ee}|$ as a function of the lightest neutrino mass, for both normal and inverted hierarchy. The filled regions correspond to the possible values of $|m_{ee}|$ in the limit of exact tri-bimaximal mixing, with mass differences computed at the central values in eq.(\ref{condnorm}). The horizontal and vertical lines are the future expected bounds on $|m_{ee}|$ and $|m_{lightest}|$ from the CUORE and KATRIN experiments, respectively.}
\end{figure}

Finally, we present in Fig.(\ref{mee}) the predictions for the values of the effective mass $|m_{ee}|$ as a function of the lightest neutrino mass, for both normal and inverted hierarchy. We also show the future sensitivity of the KATRIN   (vertical solid line) and of  CUORE \cite{Giuliani:2008zz} (horizontal solid line at 15 meV) experiments.
The main feature of the analysis is that a large set of points falls into the region of $|m_{lightest}|$ around $10^{-2}$ eV but, given the still large number of parameters of the model, many values of $|m_{ee}|$ can be obtained for both hierarchies among the experimental allowed ranges. 
On the other hand, the region above this value (the regime of degenerate spectra)
is strongly disfavoured.

\subsection{Leptogenesis}
The formal description of the asymmetry parameters can be done in the general context of arbitrary Dirac mass matrix because the final expressions are quite compact and transparent. 
The asymmetry parameters are defined as follows:
\bea
\label{epslepto}
\epsilon_i&=&\frac{1}{8\pi (\hat Y \hat Y^\dagger)_{ii}}\,\sum_{j\ne i} Im\left\{\left[(\hat Y \hat Y^\dagger)_{ij}\right]^2\right\}\, f\left(\frac{|M_j|^2}{|M_i|^2}\right)
\eea
where the {\it hat} matrices are Yukawa matrices evaluated in the basis in which the Majorana mass matrix is diagonal and $M_i$ are the Majorana masses.
For supersymmetric theories, the $f$-function is given by:
\bea
\label{eq:effe}
f(x)=-\sqrt{x}\left[\frac{2}{x-1}+\log \left(\frac{1+x}{x}\right)\right].
\eea 
Defining $\Omega$ as the unitary matrix which diagonalizes the Majorana mass matrix,  
the LO   Yukawa matrix in this basis is given by 
\bea
\nn
 v_u\,\hat Y=v_u\,\Omega^T\,Y_\nu = \Omega^T\, m_D
 \eea
and the product $\hat Y \hat Y^\dagger$ reads:
\be
 \hat Y \hat Y^\dagger = \Omega^T\, Y_\nu  Y_\nu^\dagger\, \Omega^*~.
\ee
At LO, $\Omega=U_{TBM}$   and the product $\hat Y \hat Y^\dagger$ is a diagonal matrix: 
the $\epsilon_i$ parameters are all vanishing \cite{Jenkins:2008rb}. At the next to leading order, one has to take into account the corrections to the Yukawa matrix as well as to the Majorana mass matrix, which reflects in a different structure of both the $\Omega$ and $Y_\nu$ matrices in such a way that: 
\bea
v_u\,Y_\nu&=&  \left(m_D+ v_u\,\left(\frac{\varepsilon}{\varepsilon'}\right)\,\varepsilon^2\,y_{\nu_1}\,\hat S\right) = v_u\,(Y_{LO}+\delta Y) \label{corrdir} \\
\Omega&=&U_{TBM}\,U_\phi+\delta \Omega \nn
\eea
where  $\delta \Omega$ has a structure similar to the corrections $\delta U$ computed in Sect.(\ref{subcorr})
and $\delta Y$ is of ${\cal O}(\lambda_C^4)$  compared  to its leading order result.
This means that the correction to the matrix product $\hat Y \hat Y^\dagger$ is given by:
\bea
\delta (\hat Y \hat Y^\dagger) =   (\delta \Omega)^T\, Y_{LO}  Y_{LO} ^\dagger\, \Omega^* +
\Omega^T\, Y_{LO}  Y_{LO}^\dagger\, (\delta \Omega)^* +  \Omega^T\, \delta(Y_\nu  Y_\nu^\dagger)\, \Omega^* ~.\nn
\eea
The first two terms do not contribute to the  $\epsilon_i$ parameters because they are complex conjugate of each other and do not contain any imaginary part; then the only contribution arises from the last term.
In the basis in which the charged leptons are diagonal and considering $\varepsilon$ as a real variable for simplicity, one easily obtains:
\bea
\delta(\hat Y \hat Y^\dagger) = 
\left(
\begin{array}{ccc}
\sigma_1 \,\varepsilon^{3}  & \sqrt{2}\,e^{i(\phi_1-\phi_2)}\,\sigma_2 \,\varepsilon^{3}  & 0\\
\sqrt{2}\,e^{-i(\phi_1-\phi_2)}\,\sigma_3  \,\varepsilon^{3} &  0  & 0 \\
0  &   0   &  \sigma_4 \,\varepsilon^{3}  
\end{array}
\right) 
\eea
where the $\sigma_i$ coefficients are complicated functions of $y_{\nu_i}$, $A_S$ and $B_S$. 
Then, at leading order, the $\epsilon$ parameters are given by:
\bea
\epsilon_1 &=&\left(\frac{\varepsilon}{\varepsilon'}\right)^2\,\varepsilon^4\,\frac{(A_S-B_S)^2}{4\pi} \frac{\left[y_{\nu_1}(y_{\nu_2}+2 y_{\nu_3})+3|y_{\nu_1}|^2\right]^2\,f\left(\frac{|M_2|^2}{|M_1|^2}\right)\,\sin\left[2(\phi_1-\phi_2)\right]}
{9 |y_{\nu_1}|^2+|y_{\nu_2}|^2+|y_{\nu_3}|^2-2(y_{\nu_2}y_{\nu_3})+3y_{\nu_1}y_{\nu_2}-3
y_{\nu_1}y_{\nu_3}}\, 
 \nn \\ \nn \\ \label{leptoparam}
\epsilon_2 &=& \left(\frac{\varepsilon}{\varepsilon'}\right)^2\,\varepsilon^4\,\frac{(A_S-B_S)^2}{4\pi} \frac{\left[y_{\nu_1}(y_{\nu_2}+2 y_{\nu_3})+3|y_{\nu_1}|^2\right]^2\,f\left(\frac{|M_1|^2}{|M_2|^2}\right)\,\sin\left[2(\phi_1-\phi_2)\right]}
{4|y_{\nu_2}|^2+|y_{\nu_3}|^2+4(y_{\nu_2}y_{\nu_3})}\\ \nn\\ \nn
\epsilon_3 &=&0.  
\eea
A relevant feature of the model is that $\epsilon_{3}$ always vanishes; however,
we can see that, barring possible fine-tunings in the parameters and/or suppressions or enhancements due to $f(|x|^2)\,\sin\left[2 \Delta\phi\right]$,
$\epsilon_{1,2}$  can be of the right order of magnitude to fulfill the experimental requirements for a successful leptogenesis because 
\bea
\epsilon_{1,2} \sim \lambda_C^8 \sim 6 \times 10^{-6}\,.
\eea

\section{The quark sector}
\label{quarks}
It is well known that the extension of a flavour symmetry from the neutrino sector to the quark one is highly non trivial due to the ``non-trigonometric'' structure of the quark mixing matrix $V_{CKM}$.
The common  way to introduce quarks in these kind of models is to put the two $SU(2)$ doublets of left-handed quarks  into a doublet representation of $S_4$ and assign the heaviest ones (top and bottom quarks) into singlets, in such a way to easily maintain the hierarchy among mass eigenstates. Better results are obtained if other extra symmetries, i.e.  Froggatt-Nielsen or extra $Z_N$, are introduced in order to further suppress the unwanted couplings; this is particularly true for the $m_{up}/m_{top}$ mass ratio, which otherwise tends to be larger than the experimental counterpart. It is clear that such scenarios are more flexible than the simple $S_4 \times Z_5$ illustrated in this paper; however, it is important to stress  that it is still possible to get a satisfactory description of the quark sector without invoking any other extra symmetries and  using the triplet representation of $S_4$ instead of the doublet ones. 
In the down-quark sector the simplest choice is to {\it copy} the coupling allowed in the charged lepton sector because the hierarchy between the $d$, $s$ and $b$ quark masses is quite similar to that existing for $e$, $\mu$ and $\tau$ leptons. On the other hand, the mass hierarchy in the up-quark sector does not follow such a simple prescription and we can arrange the $S_4 \times Z_5$ charge assignment in several ways. One possibility is summarized in Tab.(\ref{quarks1}), where $Q$ is a triplet of $SU(2)$ left-handed doublets.
\begin{table}[h!]
\centering
\begin{tabular}{|c||c|c|c|c|c|c|c|}
\hline
{\tt Field}& $Q$ &$ d^c$ & $s^c$ & $b^c$ & $u^c$ & $c^c$ & $t^c$\\
\hline
$S_4$ & $3_1$ &$1_2$ & $1_1$ & $1_1$ & $1_2$ & $1_1$ &$1_1$ \\
\hline
$Z_5$ & 1 & $\omega^3$ & $\omega^2$ & $\omega$ & $\omega^4$& 1 &  $\omega^4$\\
\hline
$U(1)_R$ & $1$ & $1$ & $1$ & $1$ & $1$ & $1$ & $1$\\
\hline
\end{tabular}
\caption{\it Transformation properties of quarks under $S_4 \times Z_5$ and $U(1)_R$~.}
\label{quarks1}
\end{table}
The corresponding leading order lagrangian is the following:
\bea
{\cal L}&=&\frac{y_b}{\Lambda} b^c (Q \varphi_T) \, h_d +\nn \\
&&\frac{y_{s_1}}{\Lambda^2} s^c (Q\, \varphi_T \,\varphi_T) \, h_d +
\frac{y_{s_2}}{\Lambda^2} s^c Q\, (\eta\,\varphi_T) \, h_d + \nn \\
&&\nn \frac{y_{d_1}}{\Lambda^3} d^c \,Q\,[\varphi_T (\varphi_T\varphi_T)_2]_{3_2} \, h_d +
\frac{y_{d_2}}{\Lambda^3} d^c \,Q\,[\varphi_T (\varphi_T\varphi_T)_{3_1}]_{3_2} \, h_d +\\
&&\nn
\frac{y_{d_3}}{\Lambda^3} d^c \,Q\, [\eta\, (\varphi_T\varphi_T)_{3_1}] _{3_2}\, h_d 
+ \frac{y_{d_4}}{\Lambda^3} d^c \,Q\, [\varphi_T \, (\eta\eta)_{2}] _{3_2}\, h_d +\\
&&\label{lagquark}
\frac{y_{d_5}}{\Lambda^2} d^c \,Q\,  (\Delta \varphi_S)_{3_2}\, h_d\, + 
\frac{y_t}{\Lambda} t^c (Q \varphi_T) \, h_u + \\
&&\frac{y_{c_1}}{\Lambda^2} c^c (Q\, \varphi_T \,\varphi_T) \, h_u +
\frac{y_{c_2}}{\Lambda^2} c^c Q\, (\eta\,\varphi_T) \, h_u + \nn \\ \nn
&&\frac{y_{u_1}}{\Lambda^3} u^c \,Q\,[\Delta (\varphi_T\varphi_T)_{3_1}]_{3_2} \, h_u +
\frac{y_{u_2}}{\Lambda^3} u^c \,Q\,[\varphi_S (\varphi_T\varphi_T)_2]_{3_2} \, h_u +\\
&&\nn
\frac{y_{u_3}}{\Lambda^3} u^c \,Q\, [\varphi_S\, (\varphi_T\varphi_T)_{3_1}] _{3_2}\, h_u 
+ \frac{y_{u_4}}{\Lambda^3} u^c \,Q\, [\varphi_S \, (\eta\eta)_{2}] _{3_2}\, h_u +\\
&&\nn
\frac{y_{u_5}}{\Lambda^3} u^c \,Q\,  [\eta \,(\varphi_T \varphi_S)_{3_1}] _{3_2}\, h_u\,
+\frac{y_{u_6}}{\Lambda^3} u^c \,Q \,  [\eta \,(\varphi_T \varphi_S)_{3_2}] _{3_2}\, h_u \, .
\eea
This lagrangian gives rise to a diagonal mass matrix for the down-type quarks and an almost diagonal mass matrix for the up-type quarks (all the up  entries are filled with terms of the same order  $\varepsilon^2\,\varepsilon'$). This picture slightly changes  when including the next to leading order effects. As for the case of charged leptons, in the down sector the relevant NLO corrections come from the corrections to the vacuum alignment of the flavon fields, which are all of relative ${\cal O}(1/\Lambda)$ with respect to their leading order counterparts, whereas the higher order operators give corrections of the same size only for the $s$-quark, via couplings like 
\bea
\label{newc}
(\varphi_S^3),\,(\varphi_S\Delta^2),\,(\varphi_S\xi^2),\,(\varphi_S\Delta\xi)\,,
\eea
and are much more suppressed for the down and bottom quarks. Similarly, for 
the up-type quarks the mass matrix is modified by the NLO structure of vacuum alignment of the flavon fields and, for the $c$-quark, also by the same higher order operators modifying the $s$-quark entries in eq.(\ref{newc}). All in all, the mass matrices for both type of quarks are given by:
\be
m_{down}=v_d\left(
\begin{array}{ccc} a^d_1 \, \varepsilon^3 &a^d_2 \,\varepsilon\, \varepsilon^{'2}& -a^d_2 \,\varepsilon\,\varepsilon^{'2}\\
b^d_1\, \varepsilon^{'3}& b^d_2\,\varepsilon^2 & b^d_3\,\varepsilon^{'3}\\
c^d_1\,\frac{\varepsilon^{'3}}{\varepsilon}  & c^d_2\,\frac{\varepsilon^{'3}}{\varepsilon}  & c^d_3\,\varepsilon
\end{array}
\right) \qquad
m_{up}=v_u\left(
\begin{array}{ccc} a^u_1 \, \varepsilon^2\varepsilon' &a^u_2 \,\varepsilon^2\varepsilon'& a^u_3 \,\varepsilon^2\varepsilon'\\
b^u_1\, \varepsilon^{'3}& b^u_2\,\varepsilon^2 & b^u_3\,\varepsilon^{'3}\\
c^u_1\,\frac{\varepsilon^{'3}}{\varepsilon}  & c^u_2\,\frac{\varepsilon^{'3}}{\varepsilon}  & c^u_3\,\varepsilon
\end{array}
\right)
\label{mquarkfin}
\ee
with the following mass eigenvalues:
\bea
m_d &=& v_d\,\left[a_1^d\, \varepsilon ^3 +\left(\frac{c_1^d}{c_3^d}-\frac{b_1^d}{b_3^d}\right)\,a_2^d \,\varepsilon\,\varepsilon^{'3}\right] \nn \\
m_s &=& v_d\,b_2^d\,\varepsilon^2 + {\cal O}(\varepsilon^{4})  \\
m_b &=& v_d\,c_3^d\,\varepsilon+ {\cal O}(\varepsilon^{4}) \nn
\eea
and 
\bea
m_u &=& v_u\,\left[a_1^u\, \varepsilon^2\,\varepsilon' +\left(\frac{b_1^u\,a_2^u}{b_2^u}-\frac{a_3^u\,c_1^u}{c_3^u}\right)\,\varepsilon^{'4}\right] \nn \\
m_c &=& v_u\,b_2^u\,\varepsilon^2 + {\cal O}(\varepsilon^{'4}) \\
m_t &=& v_u\,c_3^u\,\varepsilon+ {\cal O}(\varepsilon^{'4})\,. \nn
\eea
The previous mass matrices can be diagonalized by unitary matrices is such a way that:
\bea
(m_u^\dagger \, m_u)_{\text{diag}} &=& U^\dagger_{u_L}\, (m_u^\dagger \, m_u)\,U_{u_L} \nn \\
(m_d^\dagger \, m_d)_{\text{diag}} &=& U^\dagger_{d_L}\, (m_d^\dagger \, m_d)\,U_{d_L} \nn \,,
\eea
where:
\bea
U_{d_L} = \left(
\begin{array}{ccc} 1 & \left(\frac{b_1^d}{b_2^d} \,\frac{\varepsilon^{'3}}{\varepsilon^2}\right)^*& \left(\frac{c_1^d}{c_3^d} \,\frac{\varepsilon^{'3}}{\varepsilon^2}\right)^* \\
-\frac{b_1^d}{b_2^d} \,\frac{\varepsilon^{'3}}{\varepsilon^2}  & 1 & \left(\frac{c_2^d}{c_3^d} \,\frac{\varepsilon^{'3}}{\varepsilon^2}\right)^* \\
- \frac{c_1^d}{c_3^d} \,\frac{\varepsilon^{'3}}{\varepsilon^2}   & - \frac{c_2^d}{c_3^d} \,\frac{\varepsilon^{'3}}{\varepsilon^2}   & 1
\end{array}
\right) ,\qquad
U_{u_L} = \left(
\begin{array}{ccc} 1 & \left(\frac{b_1^u}{b_2^u} \,\frac{\varepsilon^{'3}}{\varepsilon^2}\right)^*& \left(\frac{c_1^u}{c_3^u} \,\frac{\varepsilon^{'3}}{\varepsilon^2}\right)^* \\
-\frac{b_1^u}{b_2^u} \,\frac{\varepsilon^{'3}}{\varepsilon^2}  & 1 & \left(\frac{c_2^u}{c_3^u} \,\frac{\varepsilon^{'3}}{\varepsilon^2}\right)^* \\
- \frac{c_1^u}{c_3^u} \,\frac{\varepsilon^{'3}}{\varepsilon^2}   & - \frac{c_2^u}{c_3^u} \,\frac{\varepsilon^{'3}}{\varepsilon^2}   & 1
\end{array}
\right)
\eea
The resulting $V_{CKM}$ is given by:
\bea
V_{CKM} = U^\dagger_{u_L}\,U_{d_L}=\left(
\begin{array}{ccc} 1 & \left[\left(\frac{b_1^d}{b_2^d} -\frac{b_1^u}{b_2^u}\right) \,\frac{\varepsilon^{'3}}{\varepsilon^2}\right]^*& 
\left[\left(\frac{c_1^d}{c_3^d} -\frac{c_1^u}{c_3^u}\right) \,\frac{\varepsilon^{'3}}{\varepsilon^2}\right]^* \\
 - \left(\frac{b_1^d}{b_2^d} -\frac{b_1^u}{b_2^u}\right) \,\frac{\varepsilon^{'3}}{\varepsilon^2} & 1 & \left[\left(\frac{c_2^d}{c_3^d}
 -\frac{c_2^u}{c_3^u}\right) \,\frac{\varepsilon^{'3}}{\varepsilon^2}\right]^* \\
 -\left(\frac{c_1^d}{c_3^d} -\frac{c_1^u}{c_3^u}\right) \,\frac{\varepsilon^{'3}}{\varepsilon^2}   & -\left(\frac{c_2^d}{c_3^d} -\frac{c_2^u}{c_3^u}\right)     
\,\frac{\varepsilon^{'3}}{\varepsilon^2}  & 1
\end{array}
\right)\,.
\eea
As for other flavour models, the matching of the $V_{CKM}$ and the quark mass ratios to their experimental values requires some fine-tunings between the Yukawas.
As anticipated, all the experimental mass ratios in the down sector are easily reproduced for the natural values $a_1^d,b_2^d,c_3^d \sim {\cal O}(1)$ because $m_d/m_s \sim m_s/m_b \sim \varepsilon \sim \lambda_C^2 $. A moderate hierarchy  is also present in the up sector and it mainly depends on the parameter $\varepsilon'$, which we estimated in Sect.(\ref{subcorr2}) to be not larger than some units of $\lambda_C^2$. This same parameter appears in the off-diagonal entries of $V_{CKM}$ with the same power in each entries. This is essentially the reason why it is difficult to  explain at the same time the mass hierarchy in the up-type quark sector and the off-diagonal values of the quark mixing matrix. For example, one can treat $\varepsilon'$ and $\tan \beta$ as free parameters of the model. 
In that case, one can use tha ratio $m_u/m_c$ to fix the value  of  $\varepsilon'$ and $m_u/m_d$ for that of $\tan \beta$; assuming for all Yukawas the natural ${\cal O}(1)$ value, we get:
\bea
\frac{m_u}{m_c} = \varepsilon' \qquad \frac{m_u}{m_d} = \left(\frac{\varepsilon'}{\varepsilon}\right)\tan \beta 
\eea
and then
\bea
\varepsilon'=\left(\frac{m_u}{m_c}\right)_{\text{exp}} \qquad \tan \beta = \lambda_C^2\,\left(\frac{m_u}{m_d}\right)_{\text{exp}}/ \left(\frac{m_u}{m_c}\right)_{\text{exp}} \sim 12\,.
\eea
It is easy to verify that the previous results are enough to accomodate all the  independent  mass ratios that can be built from six different quarks. However, $\varepsilon'$ turns out to be very small and, having assumed all Yukawas of ${\cal O}(1)$, it is really difficult to enhance the off-diagonal elements of $V_{CKM}$ to their experimental values. 
The other possibility is to preserve $\varepsilon'\sim {\cal O}(\lambda_C^2)$, requiring that
\bea
\left[\left(\frac{b_1^d}{b_2^d} -\frac{b_1^u}{b_2^u}\right)\right] \sim {\cal O} (1/\lambda_C) \qquad 
\left[\left(\frac{c_1^d}{c_3^d} -\frac{c_1^u}{c_3^u}\right) \right]\sim {\cal O} (\lambda_C) \qquad 
\left[\left(\frac{c_2^d}{c_3^d} -\frac{c_2^u}{c_3^u}\right) \,\right] \sim {\cal O} (1) \,.
\eea
This fine-tuning  is a condition on the individual Yukawas, which can be chosen to be:
\bea
c_2^d &\sim& b_1^u \sim {\cal O} (1) \nn \\
c_1^d &\sim& {\cal O} (\lambda_C) \\
b_1^d &\sim&  {\cal O} (1/\lambda_C)\,, \nn
\eea
providing that $b_2^u,c_3^u \gg 1$ to fit the charm and top quark masses. Although this situation is not completely satisfactory, it 
illustrates a way to account for many experimental informations in the quark sector using a simple and unconstrained $S_4 \times Z_5$ model, the prize to be paid being a fine-tuning in four of the Yukawa couplings appearing in the lagrangian.

Notice that it is not more difficult to modify the charge assignment proposed in Tab.(\ref{quarks1}) in such a way to obtain at 
LO a $V_{CKM}$ matrix with entries of ${\cal O}(1)$ in the Cabibbo $2\times2$ submatrix. This can be accomplished slightly changing only the $Z_5$ charge of the quark $c$. In fact, giving the assignment $c \sim (1_1,\omega^3)$ under $(S_4,Z_5)$, the $c$-quark part of lagrangian is now:
\bea
\hspace{-0.3cm}
{\cal L}_c&=&\frac{y_{c_1}}{\Lambda^2} c^c (Q\, \varphi_T \,\Delta) \, h_u +
\frac{y_{c_2}}{\Lambda^2} c^c Q\, (\varphi_T\,\varphi_S) \, h_u + 
\frac{y_{c_3}}{\Lambda^2} c^c Q\, \varphi_T \,\xi \, h_u +
\frac{y_{c_4}}{\Lambda^2} c^c Q\, (\varphi_S\,\eta) \, h_u 
\eea 
and the $c$-quark entries in the mass matrix are modified according to 
\be
(c^c,u): b^u_1\, \varepsilon\,\varepsilon' \qquad  (c^c,c): b^u_2\,\varepsilon\,\varepsilon'   \qquad  (c^c,t): b^u_3\,\varepsilon\,\varepsilon'\,.
\label{mquarkfin2}
\ee 
Correspondingly, the $u$ and $c$-quark masses are: 
\bea
m_u = v_u\,\left(a_1^u-\frac{b_1^u\,a_2^u}{b_2^u}\right)\varepsilon^2\,\varepsilon' \qquad m_c= v_u\, b^u_2\,\varepsilon\,\varepsilon'
\eea 
and the $V_{CKM}$ has the following structure:
\bea
V_{CKM} = \left(
\begin{array}{ccc}  -\frac{b_2^u}{K}-\frac{b_1^u b_1^d}{b_2^d K} \frac{\varepsilon^{'3}}{\varepsilon^2}& \frac{b_1^u}{K}-\frac{b_2^u b_1^d}{b_2^d K} \frac{\varepsilon^{'3}}{\varepsilon^2} & \frac{c_3^d (b_2^u c_1^u -b_1^u c_2^u)-c_3^u (b_2^u c_1^d -b_1^u c_2^d)}{c_3^d c_3^u K}\, \frac{\varepsilon^{'3}}{\varepsilon^2}\\
-\frac{b_1^u}{K}+\frac{b_2^u b_1^d}{b_2^d K} \frac{\varepsilon^{'3}}{\varepsilon^2}  &  -\frac{b_2^u}{K} -\frac{b_1^u b_1^d}{b_2^d K} \frac{\varepsilon^{'3}}{\varepsilon^2}&  \frac{c_3^d (b_1^u c_1^u +b_2^u c_2^u)-c_3^u (b_1^u c_1^d +b_2^u c_2^d)}{c_3^d c_3^u K}\, \frac{\varepsilon^{'3}}{\varepsilon^2}\\
 \left(-\frac{c_1^d}{c_3^d}+ \frac{c_1^u}{c_3^u}\right)\frac{\varepsilon^{'3}}{\varepsilon^2}  & \left(-\frac{c_2^d}{c_3^d}+
 \frac{c_2^u}{c_3^u}\right)\frac{\varepsilon^{'3}}{\varepsilon^2}   & 1
\end{array}
\right)\,
\eea
where, for simplicity, we introduced the short-hand notation $K=\sqrt{(b_1^{u})^2+(b_2^{u})^2}$ and considered real vevs and Yukawa couplings. In this case, one can reproduce at the same time the correct values for the (11) and (22) entries of $V_{CKM}$ as well as the mass ratio $m_u/m_c$ imposing $b_2^u \sim {\cal O}(\lambda_C^2)$; playing with $\varepsilon'$ and the other Yukawas one can also reproduce the off-diagonal entries, but one or more fine-tunings in the coefficients in front of $\varepsilon^{'3}/\varepsilon^2$ ratio are obviously needed.
\section{Conclusions}
\label{concl}
We have presented and discussed an $S_4$ model for TB mixing (of the see-saw type) and quark mixing which, in
spite of being based on a most economical flavour symmetry and field content, it is still phenomenologically
viable. In the neutrino sector, we realized the most general neutrino mass matrix diagonalized at LO by TB mixing. At the NLO,  all mixing angles receive corrections at the level of 
$\mathcal{O}(\lambda_C^2)$ and, in particular, the elusive $\theta_{13}$ is predicted to be  within the sensitivity of the experiments which are now in preparation and will take data in the near future. Also, we have shown that the leptogenesis parameters $\epsilon_{1,2}$ can be of the right order of magnitude to satisfy the requirements needed to reproduce the observed asymmetry and that an acceptable phenomenology is still obtainable even with all Yukawa couplings at ${\cal O}(1)$.

In the charged lepton sector, the mass hierarchy is determined by the $S_4\times Z_5$ flavour symmetry itself without invoking a Froggatt-Nielsen $U(1)$ symmetry. The NLO corrections to the mass matrix turn out to be relevant for a $\theta_{13}\ne 0$.

In the quark sector, we have discussed in detail to which extent the model can reproduce the data on masses and mixing, emphasizing  where it is successful and  the reasons for its failures. Two different LO examples for $V_{CKM}$ have been given, one proportional to the identity matrix and the other with ${\cal O}(1)$ elements in the $2\times2$ sector; both realizations need almost the same amount of fine-tuning  to reproduce the off-diagonal entries at the NLO.   

\section{Acknowledgements}
I am grateful to Guido Altarelli for very stimulating discussions. I also wish to thank Luca Merlo for some interesting comments.
This work was supported by the Deutsche Forschungs-gemeinschaft, contract WI 2639/2-1. 
\appendix
\section{The Group $S_4$}
We adopt the following convention for the generators $S$ and $T$, according to \cite{bazz}
\be
S^4= T^3=  (ST^2)^2=\unity
\ee
In the different representations, they can be written as reported in Tab.(\ref{tabrepre}):

\begin{table}[h!]
\centering
\begin{tabular}{|c||c|c|c|c|c|}
\hline
{\tt rep}& $1_1$ &$1_2$ & $2$ & $3_1$ & $3_2$ \\   \hline   \hline   
$S$  & 1  &  -1 & $\left(\begin{array}{cc} 0 & 1 \\ 1 & 0 \\ \end{array} \right)$  & $\dfrac{1}{3}\left(
                                 \begin{array}{ccc}
                                   -1 & 2\om & 2\om^2 \\
                                   2\om & 2\om^2 & -1 \\
                                   2\om^2 & -1 & 2\om \\
                                 \end{array}
                               \right)$  & $\dfrac{1}{3}\left(
                                 \begin{array}{ccc}
                                   1 & -2\om & -2\om^2 \\
                                   -2\om & -2\om^2 & 1 \\
                                   -2\om^2 & 1 & -2\om \\
                                 \end{array}
                               \right)$ \\
\hline
$T$ & 1  & 1 & $\left(\begin{array}{cc} \om & 0 \\ 0 & \om^2 \\ \end{array} \right)$  &  $\left(
                                              \begin{array}{ccc}
                                                1 & 0 & 0 \\
                                                0 & \om^2 & 0 \\
                                                0 & 0 & \om \\
                                              \end{array}
                                            \right)$ &$\left(
                                              \begin{array}{ccc}
                                                1 & 0 & 0 \\
                                                0 & \om^2 & 0 \\
                                                0 & 0 & \om \\
                                              \end{array}
                                            \right)$ \\
\hline
\end{tabular}
\caption{\it Generators $S$ and $T$ in different representations.}
\label{tabrepre}
\end{table}

The 24 elements of the group belong to five conjugacy classes
\begin{eqnarray*}
&&{\cal C}_1:1\\
&&{\cal C}_2:\;S^2,\;TS^2T^2,\;S^2TS^2T^2\\
&&{\cal C}_3:\;T,\;T^2,\;S^2T,\;S^2T^2,\;STST^2\\
&&~~~~~~STS,\;TS^2,\;T^2S^2\\
&&{\cal
C}_4:\;ST^2,\;T^2S,\;TST\\
&&~~~~~~TSTS^2,\;STS^2,\;S^2TS\\
&&{\cal
C}_5:\;S,\;TST^2,\;ST,\;\;TS,\;S^3,\;S^3T^2\,.
\end{eqnarray*}

The explicit expression of the elements in the 2-dimensional representation is:
\begin{itemize}
\item $\cC_{1,2}:$ \qquad $\left(
                          \begin{array}{cc}
                            1 & 0 \\
                            0 & 1 \\
                          \end{array}
                        \right)$
\item$\cC_3:$ \qquad  $\left(
                      \begin{array}{cc}
                        \om & 0 \\
                        0 & \om^2 \\
                      \end{array}
                    \right)\;,\;
                    \left(
                      \begin{array}{cc}
                        \om^2 & 0 \\
                        0 & \om \\
                      \end{array}
                    \right)$
\item$\cC_{4,5}:$ \qquad   $\left(
                      \begin{array}{cc}
                        0 & 1 \\
                        1 & 0 \\
                      \end{array}
                    \right)\;,\;
                    \left(
                      \begin{array}{cc}
                        0 & \om \\
                        \om^2 & 0 \\
                      \end{array}
                    \right)\;,\;
                    \left(
                      \begin{array}{cc}
                        0 & \om^2 \\
                        \om & 0 \\
                      \end{array}
                    \right)\;,$
\end{itemize}

while for the 3-dimensional representation $3_1$ the elements are

\begin{itemize}
\item$\cC_1:$  \qquad $I=\left(
                 \begin{array}{ccc}
                   1 & 0 & 0 \\
                   0 & 1 & 0 \\
                   0 & 0 & 1 \\
                 \end{array}
               \right)$ 
\item$\cC_2:$ \qquad  $S^2$=$\dfrac{1}{3}\left(
                 \begin{array}{ccc}
                   -1 & 2 & 2 \\
                   2 & -1 & 2 \\
                   2 & 2 & -1 \\
                 \end{array}
               \right)$ \qquad
            $TS^2T^2$=   $\dfrac{1}{3}\left(
                 \begin{array}{ccc}
                   -1 & 2\om & 2\om^2 \\
                   2\om^2 & -1 & 2\om \\
                   2\om & 2\om^2 & -1 \\
                 \end{array}
               \right)$ 
\item[{}] \qquad
           $S^2TS^2T^2$ =   $\dfrac{1}{3}\left(
                 \begin{array}{ccc}
                   -1 & 2\om^2 & 2\om \\
                   2\om & -1 & 2\om^2 \\
                   2\om^2 & 2\om & -1 \\
                 \end{array}
               \right) $
\item$\cC_3:$\qquad  $T=\left(
                 \begin{array}{ccc}
                   1 & 0 & 0 \\
                   0 & \om^2 & 0 \\
                   0 & 0 & \om \\
                 \end{array}
               \right)$ \qquad $T^2=   \left(
                 \begin{array}{ccc}
                   1 & 0 & 0 \\
                   0 & \om & 0 \\
                   0 & 0 & \om^2 \\
                 \end{array}
               \right)$\qquad $S^2T=\dfrac{1}{3}\left(
                 \begin{array}{ccc}
                   -1 & 2\om^2 & 2\om \\
                   2 & -\om^2 & 2\om \\
                   2 & 2\om^2 & -\om \\
                 \end{array}
               \right)$
\item[{}]\qquad $S^2T^2=   \dfrac{1}{3}\left(
                 \begin{array}{ccc}
                   -1 & 2\om & 2\om^2 \\
                   2 & -\om & 2\om^2 \\
                   2 & 2\om & -\om^2 \\
                 \end{array}
               \right)$ \qquad $STST^2=  \dfrac{1}{3}\left(
                 \begin{array}{ccc}
                   -1 & 2 & 2 \\
                   2\om^2 & -\om^2 & 2\om^2 \\
                   2\om & 2\om & -\om \\
                 \end{array}
               \right)$
\item[{}]\qquad$STS=\dfrac{1}{3}\left(
                 \begin{array}{ccc}
                   -1 & 2\om^2 & 2\om \\
                   2\om^2 & -\om & 2 \\
                   2\om & 2 & -\om^2 \\
                 \end{array}
               \right)$  \qquad
 $TS^2=\dfrac{1}{3}\left(
                 \begin{array}{ccc}
                   -1 & 2\om & 2\om^2 \\
                   2\om & -\om^2 & 2 \\
                   2\om^2 & 2 & -\om \\
                 \end{array}
               \right)$ 
\item[{}]\qquad
$T^2S^2=   \dfrac{1}{3}\left(
                 \begin{array}{ccc}
                   -1 & 2 & 2 \\
                   2\om & -\om & 2\om \\
                   2\om^2 & 2\om^2 & -\om^2 \\
                 \end{array}
               \right)$
\item$\cC_4:$ \qquad $ST^2=\dfrac{1}{3}\left(
                 \begin{array}{ccc}
                   -1 & 2\om^2 & 2\om \\
                   2\om & 2 & -\om^2 \\
                   2\om^2 & -\om & 2 \\
                 \end{array}
               \right)$\qquad
           $T^2S=    \dfrac{1}{3}\left(
                 \begin{array}{ccc}
                   -1 & 2\om & 2\om^2 \\
                   2\om^2 & 2 & -\om \\
                   2\om & -\om^2 & 2 \\
                 \end{array}
               \right)$ 
\item[{}]\qquad$TST=    \dfrac{1}{3}\left(
                 \begin{array}{ccc}
                   -1 & 2 & 2 \\
                   2 & 2 & -1 \\
                   2 & -1 & 2 \\
                 \end{array}
               \right)$  \qquad $TSTS^2=\left(
                 \begin{array}{ccc}
                   1 & 0 & 0 \\
                   0 & 0 & 1 \\
                   0 & 1 & 0 \\
                 \end{array}
               \right)$ 
\item[{}]\qquad$STS^2=
               \left(
                 \begin{array}{ccc}
                   1 & 0 & 0 \\
                   0 & 0 & \om \\
                   0 & \om^2 & 0 \\
                 \end{array}
               \right) $
\qquad$S^2TS=  \left(
                 \begin{array}{ccc}
                   1 & 0 & 0 \\
                   0 & 0 & \om^2 \\
                   0 & \om & 0 \\
                 \end{array}
               \right)$
\item$\cC_5:$ \qquad $S=\dfrac{1}{3}\left(
                 \begin{array}{ccc}
                   -1 & 2\om & 2\om^2 \\
                   2\om & 2\om^2 & -1 \\
                   2\om^2 & -1 & 2\om \\
                 \end{array}
               \right)$\qquad 
            $TST^2=  \dfrac{1}{3}\left(
                 \begin{array}{ccc}
                   -1 & 2\om^2 & 2\om \\
                   2 & 2\om^2 & -\om \\
                   2 & -\om^2 & 2\om \\
                 \end{array}
               \right)$
\item[{}]\qquad$ST=  \dfrac{1}{3}\left(
                 \begin{array}{ccc}
                   -1 & 2 & 2 \\
                   2\om & 2\om & -\om \\
                   2\om^2 & -\om^2 & 2\om^2 \\
                 \end{array}
               \right)$  \qquad $TS=\dfrac{1}{3}\left(
                 \begin{array}{ccc}
                   -1 & 2\om & 2\om^2 \\
                   2 & 2\om & -\om^2 \\
                   2 & -\om & 2\om^2 \\
                 \end{array}
               \right)$
\item[{}]\qquad $S^3=    \dfrac{1}{3}\left(
                 \begin{array}{ccc}
                   -1 & 2\om^2 & 2\om \\
                   2\om^2 & 2\om & -1 \\
                   2\om & -1 & 2\om^2 \\
                 \end{array}
               \right)$\qquad
            $S^3T^3=   \dfrac{1}{3}\left(
                 \begin{array}{ccc}
                   -1 & 2 & 2 \\
                   2\om^2 & 2\om^2 & -\om^2 \\
                   2\om & -\om & 2\om \\
                 \end{array}
               \right)$ 
\end{itemize}
For the 3-dimensional representation $3_2$, the matrices representing the elements of the group can be 
obtained from the list for the representation $3_1$ in the following way: for $\cC_{1,2,3}$ are the same, while for $\cC_{4,5}$ are the opposite. 
\\
In the previous basis, the Clebsch-Gordan coefficients are as follows
($\alpha_i$ indicates the elements of the first
representation of the product and $\beta_i$ the second one):

\[
\begin{array}{lcl}
1_1\otimes\eta&=&\eta\otimes1_1=\eta\quad\text{with $\eta$ any
representation}\\[-10pt]
\\[8pt]
1_2\otimes1_2&=&1_1\sim\alpha\beta\\[-10pt]
\\[8pt]
1_2\otimes2&=&2\sim\left(\begin{array}{c}
                    \alpha\beta_1 \\
                    -\alpha\beta_2 \\
            \end{array}\right)\\[-10pt]
\\[8pt]
1_2\otimes3_1&=&3_2\sim\left(\begin{array}{c}
                    \alpha\beta_1 \\
                    \alpha\beta_2 \\
                    \alpha\beta_3 \\
                    \end{array}\right)\\[-10pt]
\\[8pt]
1_2\otimes3_2&=&3_1\sim\left(\begin{array}{c}
                            \alpha\beta_1 \\
                            \alpha\beta_2 \\
                            \alpha\beta_3 \\
                    \end{array}\right) \,.
\end{array}
\]
The multiplication rules with the 2-dimensional
representation are the following:
\[
\begin{array}{ll}
2\otimes2=1_1\oplus1_2\oplus2&\quad
\text{with}\quad\left\{\begin{array}{l}
                    1_1\sim\alpha_1\beta_2+\alpha_2\beta_1\\[-10pt]
                    \\[8pt]
                    1_2\sim\alpha_1\beta_2-\alpha_2\beta_1\\[-10pt]
                    \\[8pt]
                    2\sim\left(\begin{array}{c}
                        \alpha_2\beta_2 \\
                        \alpha_1\beta_1 \\
                    \end{array}\right)
                    \end{array}
            \right.\\[-10pt]
\\[8pt]
2\otimes3_1=3_1\oplus3_2&\quad
\text{with}\quad\left\{\begin{array}{l}
                    3_1\sim\left(\begin{array}{c}
                        \alpha_1\beta_2+\alpha_2\beta_3 \\
                        \alpha_1\beta_3+\alpha_2\beta_1 \\
                        \alpha_1\beta_1+\alpha_2\beta_2 \\
                    \end{array}\right)\\[-10pt]
                    \\[8pt]
                    3_2\sim\left(\begin{array}{c}
                        \alpha_1\beta_2-\alpha_2\beta_3\\
                        \alpha_1\beta_3-\alpha_2\beta_1 \\
                        \alpha_1\beta_1-\alpha_2\beta_2 \\
                    \end{array}\right)\\
                    \end{array}
            \right.\\[-10pt]
\\[8pt]
2\otimes3_2=3_1\oplus3_2&\quad
\text{with}\quad\left\{\begin{array}{l}
                    3_1\sim\left(\begin{array}{c}
                        \alpha_1\beta_2-\alpha_2\beta_3\\
                        \alpha_1\beta_3-\alpha_2\beta_1 \\
                        \alpha_1\beta_1-\alpha_2\beta_2 \\
                    \end{array}\right)\\[-10pt]
                    \\[8pt]
                    3_2\sim\left(\begin{array}{c}
                        \alpha_1\beta_2+\alpha_2\beta_3 \\
                        \alpha_1\beta_3+\alpha_2\beta_1 \\
                        \alpha_1\beta_1+\alpha_2\beta_2 \\
                    \end{array}\right)\\
                    \end{array}
            \right.\\
\end{array}
\]
The multiplication rules with the 3-dimensional
representations are the following:
\[
\begin{array}{ll}
3_1\otimes3_1=3_2\otimes3_2=1_1\oplus2\oplus3_1\oplus3_2\qquad
\text{with}\quad\left\{
\begin{array}{l}
1_1\sim\alpha_1\beta_1+\alpha_2\beta_3+\alpha_3\beta_2 \\[-10pt]
                    \\[8pt]
2\sim\left(
     \begin{array}{c}
       \al_2\beta_2+\al_1\beta_3+\al_3\beta_1 \\
       \al_3\beta_3+\al_1\beta_2+\al_2\beta_1 \\
     \end{array}
   \right)\\[-10pt]
   \\[8pt]
3_1\sim\left(\begin{array}{c}
         2\al_1\beta_1-\alpha_2\beta_3-\alpha_3\beta_2 \\
         2\al_3\beta_3-\alpha_1\beta_2-\alpha_2\beta_1 \\
         2\al_2\beta_2-\alpha_1\beta_3-\alpha_3\beta_1 \\
        \end{array}\right)\\[-10pt]
        \\[8pt]
3_2\sim\left(\begin{array}{c}
         \alpha_2\beta_3-\alpha_3\beta_2 \\
         \alpha_1\beta_2-\alpha_2\beta_1 \\
         \alpha_3\beta_1-\alpha_1\beta_3 \\
    \end{array}\right)
\end{array}\right.
\end{array}
\]
\[
\begin{array}{ll}
3_1\otimes3_2=1_2\oplus2\oplus3_1\oplus3_2\qquad
\text{with}\quad\left\{
\begin{array}{l}
1_2\sim\alpha_1\beta_1+\alpha_2\beta_3+\alpha_3\beta_2\\[-10pt]
        \\[8pt]
2\sim\left(
     \begin{array}{c}
       \al_2\beta_2+\al_1\beta_3+\al_3\beta_1 \\
       -\al_3\beta_3-\al_1\beta_2-\al_2\beta_1 \\
     \end{array}
   \right)\\[-10pt]
        \\[8pt]
3_1\sim\left(\begin{array}{c}
         \alpha_2\beta_3-\alpha_3\beta_2 \\
         \alpha_1\beta_2-\alpha_2\beta_1 \\
         \alpha_3\beta_1-\alpha_1\beta_3 \\
    \end{array}\right)\\[-10pt]
        \\[8pt]
3_2\sim\left(\begin{array}{c}
         2\al_1\beta_1-\alpha_2\beta_3-\alpha_3\beta_2 \\
         2\al_3\beta_3-\alpha_1\beta_2-\alpha_2\beta_1 \\
         2\al_2\beta_2-\alpha_1\beta_3-\alpha_3\beta_1 \\
    \end{array}\right)\\
\end{array}\right.
\end{array}
\]

\end{document}